\documentclass[3p, 10pt]{elsarticle}

\usepackage{lineno,hyperref}
\modulolinenumbers[5]

\journal{Journal of Computational Physics}

\makeatletter
\providecommand{\doi}[1]{%
  \begingroup
    \let\bibinfo\@secondoftwo
    \urlstyle{rm}%
    \href{http://dx.doi.org/#1}{%
      doi:\discretionary{}{}{}%
      \nolinkurl{#1}%
    }%
  \endgroup
}
\makeatother

\usepackage{graphicx}	
\usepackage{float}
\usepackage{subcaption} 
\usepackage{amsmath}	
\usepackage{amssymb}	
\usepackage{multirow}   
\usepackage{upgreek}  

\usepackage{color}

\usepackage{xspace}
\newcommand{\textcode}[1]{\textsc{#1}}

\newcommand{\Athenapp}{\textcode{Athena++}\xspace}

\newcommand{\kathena}{\textcode{K-Athena}\xspace}
\newcommand{\AthenaPK}{\textcode{AthenaPK}\xspace}
\newcommand{\Parthenon}{\textcode{Parthenon}\xspace}

\begin{document}

\begin{frontmatter}

\title{A Robust, Performance-Portable Discontinuous Galerkin Method for Relativistic Hydrodynamics}

\author[snl,pamsu,cmsemsu]{Forrest W. Glines\corref{correspondingauthor}}
\cortext[correspondingauthor]{fglines@sandia.gov}

\author[snl]{Kristian R.C. Beckwith}
\author[snl]{Joshua R. Braun}
\author[snl]{Eric C. Cyr}
\author[snl]{Curtis C. Ober}
\author[nvidia]{Matthew Bettencourt}
\author[snl]{Keith L. Cartwright}
\author[snl]{Sidafa Conde}
\author[snl]{Sean T. Miller}
\author[snl]{Nicholas Roberds}
\author[snl]{Nathan V. Roberts}
\author[snl]{Matthew S. Swan}
\author[snl]{Roger Pawlowski}

\fntext[snl]{Sandia National Laboratories}
\fntext[pamsu]  {Department of Physics and Astronomy, Michigan State University}
\fntext[cmsemsu]{Department of Computational Mathematics, Science and Engineering, Michigan State University}
\fntext[nvidia]{NVIDIA Corporation}

\begin{abstract}
In this work, we present a discontinuous-Galerkin method for evolving
  relativistic hydrodynamics. We include an exploration of analytical and
  iterative methods to recover the primitive variables from the conserved
  variables for the ideal equation of state and the Taub-Matthews approximation
  to the Synge equation of state. We also present a new operator for enforcing a
  physically permissible conserved state at all basis points within an element
  while preserving the volume average of the conserved state. We implement this
  method using the Kokkos performance-portability library to enable running at
  performance on both CPUs and GPUs. We use this method to explore the
  relativistic Kelvin-Helmholtz instability compared to a finite volume method.
  Last, we explore the performance of our implementation on CPUs and GPUs.
\end{abstract}


\end{frontmatter}


\section{Introduction} \label{sec:intro}

Many high energy astrophysical and terrestrial plasmas attain relativistic
velocities and temperatures.  Examples from astrophysics include jets from active galactic nuclei
\citep{blandfordRelativisticJetsActive2019}, accretion flows onto black holes
\citep{villiersMagneticallyDrivenAccretion2003}, and gamma-ray bursts
\citep{kumarPhysicsGammarayBursts2015}. In terrestrial systems, relativistic flows can
also play a crucial role  in a broad range of accelerator systems, including
magnetically insulated transmission lines (MITLs) utilized in (for example) the Z machine
at Sandia National Laboratories \cite{sinarsReviewPulsedPowerdriven2020}.  In all of
these plasmas, velocities close to the speed of light lead to an apparent
increase of mass as measured by a stationary observer while relativistic
particle velocities at high temperatures lead to a non-linear increase in
pressure. Non-relativistic hydrodynamics are insufficient to model such flows
-- a relativistic treatment of the fluid is required.  Numerical solutions for
relativistic hydrodynamics were first pioneered in the 1960's and 1970's by
\citet{mayHydrodynamicCalculationsGeneralRelativistic1966} and
\citet{wilsonNumericalStudyFluid1972}. High-resolution shock-capturing
solutions followed suit, with an early review of those methods given by
\citet{martiNumericalHydrodynamicsSpecial2003}.

When modeling complex systems with small time step constraints, higher order
methods are advantageous for efficiently achieving high accuracy.
Discontinuous Galerkin methods have become standard in fluid dynamics for
enabling high-order methods in complex geometries.  High-order
discontinuous-Galerkin methods afford enhanced data locality when compared with
finite volume methods of similar order
\citep{fuhryDiscontinuousGalerkinMethods2014}. Given the trend in compute
performance outpacing memory performance in newer architectures such as
graphics processing units (GPUs), the higher arithmetic intensity of
discontinuous-Galerkin methods will permit higher computational efficiency due
to higher arithmetic intensity algorithms using more of the growing
computational throughput while using less of the stagnant memory bandwidth,
enabling higher fidelity simulations compared to finite volume simulations for
equivalent computational resources.

In this work, we present a robust, performance-portable discontinuous-Galerkin
method for relativistic hydrodynamics. In \S\ref{sec:srhd_equations} we present
a formulation of the equations of relativistic hydrodynamics that allows for a
range of equations of state; we present two such possibilities: (1) an ideal
equation of state, which approximates a perfect gas but assumes a constant
adiabatic index for a relativistic perfect gas, and (2) an approximation to the
Synge gas from \citet{mathewsHydromagneticFreeExpansion1971}, where the Synge
equation of state models a relativistic perfect gas
\citep{syngeRelativisticGas1957}. We discuss the discretization of the system
using a discontinuous-Galerkin technique and discuss
strong-stability-preserving time discretization techniques. To enable robust
higher order discretization, in \S\ref{sec:physicality_enforcing_operator} we present a new and novel
physicality-enforcing operator for discontinuous-Galerkin methods for
relativistic hydrodynamics. The method smooths conserved variables within
individual cells to the cell volume averages until all basis
points within the cell satisfy conditions for physicality (i.e.
positive density and pressure and flow speed under the speed of light). We
implement the method for relativistic hydrodynamics using the Kokkos
performance portability library to enable running on both CPUs and GPU
\citep{CarterEdwards20143202}.

A key part of any algorithm for relativistic hydrodynamics is the method by
which the non-linear relationship between primitive variables and the conserved
state is solved. In \S\ref{sec:conserved_to_primitive}, we compare analytical
and iterative methods for recovering the primitive variables from the conserved
variables for both equations of state, across a range of different hardware
platforms and compilers as facilitated by Kokkos, finding that for the ideal gas our
iterative method following \citet{riccardiPrimitiveVariableRecovering2008} is
faster, more robust, and more accurate than an analytical method, but the exact
reverse is true for an approximation to the Synge gas.

We proceed to validate the method using several tests (discussed in detail in
\S\ref{sec:tests}), exploring convergence of the method to analytical solutions
of relativistic linear waves, convergence to high resolution reference
solutions of a range of 1D shock tubes, evolution of 2D Riemann problems, and
growth rates of the relativistic Kelvin-Helmholtz instability with two
different initial perturbations. Using a 0th order basis, we find that the
method performs comparably to 1st order finite volume methods, as expected.
Using higher order bases we see the expected level of convergence for smooth
flows. In fluid systems with shocks, the method requires the
physicality-enforcing operator presented here and exhibits expected rates of
convergence around shocks. Additionally, with the exploration of the growth
rate of the Kelvin-Helmholtz problem, we show that using the more accurate HLLC
Riemann solver \citep{mignoneHLLCSolverRelativistic2006} instead of the HLL
solver \citep{schneiderNewAlgorithmsUltrarelativistic1993} has a greater impact
on the growth rate than basis order or resolution. We further utilize this test
problem to demonstrate a range of performance portability results in
\S\ref{sec:performance} before summarizing our results and conclusions in
\S\ref{sec:conclusions}.

\nopagebreak
\section{Theoretical Background and Discretization} \label{sec:method}

In this section, we describe our method for relativistic hydrodynamics in a
discontinuous-Galerkin code, starting by reviewing the equations for
relativistic hydrodynamics in \S\ref{sec:srhd_equations}, including a
discussion of the equation of state. Then, in \S\ref{sec:srhd_in_dg}, we give
the general discontinuous-Galerkin method for solving the relativistic
hydrodynamics equations as a set of hyperbolic equations with computation of
fluxes given in \S\ref{sec:riemann_solvers}.  Last, in
\S\ref{sec:physicality_enforcing_operator}, we present a new operator that enforces
physicality of all basis points within a cell while maintaining the volume
average within the cell.


\subsection{Special Relativistic Hydrodynamics} \label{sec:srhd_equations}

The special relativistic hydrodynamics equations for a relativistic fluid are
given by a set of hyperbolic conservation laws,
\begin{equation}
  \label{eq:srhd_conservation_laws}
  \partial_t \mathbf{U} + \nabla \cdot \mathcal{F}[\mathbf{W}(\mathbf{U})] = 0
\end{equation}
where the conserved variables $\mathbf{U} = [ D, \mathbf{M},E ]^T$ are the
relativistic density, relativistic specific momentum, and the total energy
density including energy from the rest mass. The flux is
\begin{equation}
  \label{eq:srhd_flux}
  \mathcal{F}[\mathbf{W}(\mathbf{U})] = \begin{bmatrix} 
    \rho \mathbf{u} \\ 
    \frac{\rho h}{c^2} \mathbf{u} \otimes \mathbf{u} + P \mathbf{I} \\
    \gamma \rho h \mathbf{u}
  \end{bmatrix},
\end{equation}
where the rest mass density $\rho$, the three spacelike components of the
4-velocity denoted here with $\mathbf{u}$, and the pressure $P$ comprises the
primitive state $\mathbf{W}(\mathbf{U}) = [\rho,\mathbf{u},P]^T$.  The specific
enthalpy $h$ is given by
\begin{equation}
  h = \frac{e + P}{\rho}
\end{equation}
where $e$ is the specific internal energy. The conserved state $\mathbf{U}$ can be
determined from the primitive state $\mathbf{W}$ by
\begin{equation}
  \label{eq:primitive_to_conserved}
  \mathbf{U} = \begin{bmatrix}
      \gamma \rho \\
      \gamma (e + P) \mathbf{u}/c^2 \\
      \gamma^2 ( e + P) - P
    \end{bmatrix}
    = \begin{bmatrix}
      \gamma \rho \\
      \gamma \rho h \mathbf{u}/c^2 \\
      \gamma^2 \rho h - P
    \end{bmatrix}
    \equiv \begin{bmatrix}
      D \\
      \mathbf{M}\\
      E
    \end{bmatrix}
\end{equation}
where $\gamma \equiv \sqrt{1 + |\mathbf{u}|^2/c^2}$ is the Lorentz factor and
$D$, $\mathbf{M}$, and $E$ are the relativistic density, relativistic momentum
density, and total energy density respectively. We also find it convenient to
use the three-velocity $\mathbf{v}$ at times, which relates to $\mathbf{u}$
following $\mathbf{u} = \gamma \mathbf{v}$ and the Lorentz velocity following
$\gamma = 1/\sqrt{ 1 - |\mathbf{v}|^2/c^2 }$. 

\subsection{Equations of State} \label{sec:srhd_eos}

The relativistic hydrodynamics equations in Eq.~\ref{eq:srhd_conservation_laws}
are not complete; an equation of state is used to close the system. Following
\citet{ryuEquationStateNumerical2006}, we express the equation of state by
relating $h$ to the primitive variables
\begin{equation}
  h \equiv h(\rho, P).
\end{equation}
The equation of state also determines the sound speed $c_s$, which is given by
\begin{equation}
  c_s^2 = - \frac{\rho}{nh}\frac{\partial h}{\partial \rho } \quad \text{with} \quad  n = \rho \frac{\partial h}{\partial P} -1
\end{equation}
where $n$ is the polytropic index. In this work, we explore two choices of
equation of state: the equation of state of an ideal gas and the Taub-Matthews approximation
to the Synge equation of state described in
\citet{mathewsHydromagneticFreeExpansion1971}.

In a relativistic perfect gas, the adiabatic index decreases with
temperature, starting with $\Gamma=5/3$ for non-relativistic temperatures when
$P/\rho \ll c^2$ and decreasing to $\Gamma = 4/3$ for relativistic temperatures
when $P/\rho \gg c^2$. The equation of state of the perfect gas is given by the Synge
gas \citep{syngeRelativisticGas1957} :
\begin{equation}
  h = c^2 \frac{ K_3 \left ( c^2/\Theta \right ) }{ K_2 \left ( c^2/\Theta \right ) }
\end{equation}
where $K_2$ and $K_3$ are modified Bessel functions of the second kind and
$\Theta \equiv P/\rho$ is a temperature-like variable. From a computational
standpoint, however, there are significant drawbacks, as these Bessel functions
are both expensive to compute and can introduce inaccuracy due to limited
machine precision. Even worse, the Bessel functions need to be inverted to
recover the primitive variables from conserved variables, which greatly
increases computational costs. Consequently, approximations to the equation of
state are usually used in simulations. 

The simplest approximation to the relativistic perfect gas is the ideal
equation of state, which assumes a constant adiabatic index. The enthalpy for
the ideal gas is given by
\begin{equation}
  \label{eq:id_enthalpy}
  h  = c^2 + \frac{\Gamma}{\Gamma - 1} \Theta
\end{equation}
where the constant $\Gamma$ is the adiabatic index (ratio of specific heats.)
The corresponding speed of sound is then:
\begin{equation}
  \frac{c_s^2}{c^2} = \Gamma \frac{ \Theta}{ h}.
\end{equation}
For non-relativistic temperatures when $\Theta \ll c^2$, an adiabatic index of
$\Gamma=5/3$ best approximates the perfect gas (consistent with
non-relativistic theory) while for relativistic temperatures when $\Theta \gg
c^2$ and adiabatic index of $\Gamma=4/3$ is appropriate. The ideal equation of
state is common for relativistic hydrodynamics simulations. However,
relativistic fluid systems can have relativistic and non-relativistic temperatures
simultaneously at different locations within the fluid, throwing into question
the use of a constant adiabatic index across the simulation.  Additionally,
\citet{taubRelativisticRankineHugoniotEquations1948} showed that $\Gamma \geq
4/3$ becomes inconsistent with relativistic kinetic theory as $\Theta/c^2
\rightarrow \infty$, suggesting that adiabatic indices above $4/3$ are
unphysical for ultra-relativistic temperatures.

\begin{figure}
  \centering
  \includegraphics[]{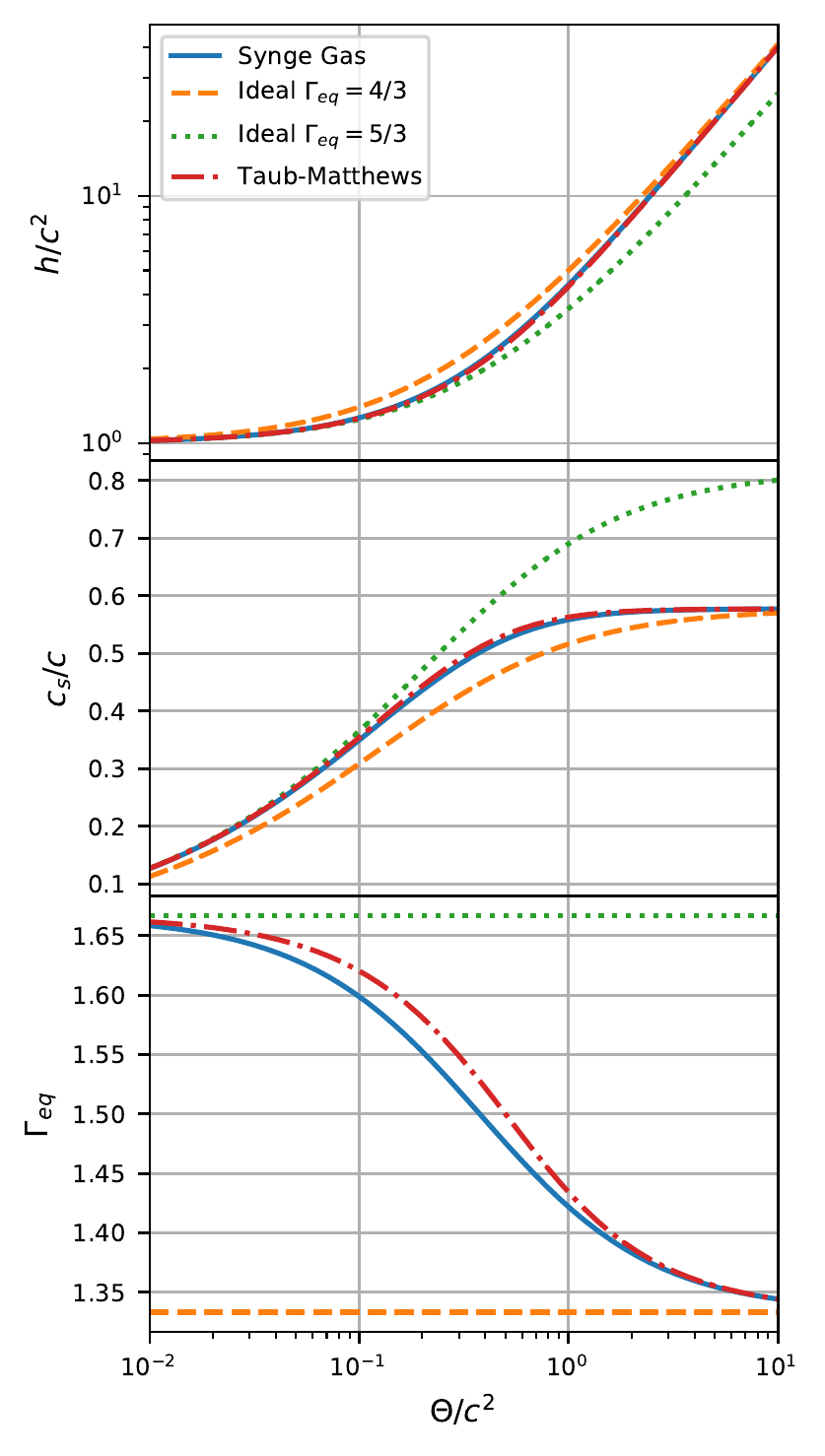}
  \caption{
    \label{fig:eos_comparison}
    Enthalpy (top), sound speed (middle), and equivalent adiabatic index
    (bottom) as a function of the temperature proxy $\Theta/c^2$ for the Synge gas
    (solid blue),  ideal equation of state with a relativistic $\Gamma=4/3$
    (dashed orange) and a non-relativistic $\Gamma=5/3$ (finely dashed green),
    and the Taub-Matthews approximation to the Synge gas (dot-dashed red). With
    the Synge and Taub-Matthews equations of state, each of the quantities
    shown here vary smoothly between the two extremes of the ideal equation of
    state as $\Theta/c^2$ changes from non-relativistic to relativistic. The
    Taub-Matthews equation of state provides a reasonable approximation to the
    Synge gas while remaining simple for computation.
  } 
\end{figure}

A more accurate approximation to the Synge gas that is still computationally
efficient is the Taub-Matthews approximation to the Synge gas, which we will
refer to as the Taub-Matthews equation of state
\citep{mathewsHydromagneticFreeExpansion1971}. In this approximation, the
enthalpy is given by:
\begin{equation}
  h  = \frac{5}{2}\Theta  + \frac{3}{2} \sqrt { \Theta^2 + \frac{4}{9}c^4 }
\end{equation}
with the corresponding sound speed:
\begin{equation}
  \frac{c_s^2}{c^2} = \frac{3 \Theta^2 + 5 \Theta \sqrt{ \Theta^2 + \frac{4}{9}c^4 } }
  { 12 \Theta^2 + 2 c^4 + 12 \Theta \sqrt{ \Theta^2 + \frac{4}{9}c^4 } }.
\end{equation}
The Taub-Matthews equation of state satisfies the conditions for causality at
high temperatures while correctly approximating the ideal gas equation of state
for a subrelativistic gas at low temperatures
\citep{mathewsHydromagneticFreeExpansion1971}. As such, the Taub-Matthews
equation of state effectively simulates an ideal gas with an adiabatic index
that varies from $\Gamma = 5/3$ as $\Gamma=4/3$ as $\Theta$ is taken from
$\Theta \rightarrow 0$ to $\Theta \rightarrow \infty$. More formally, this can
be seen through defining an equivalent adiabatic index\footnote{Note that since we have
not defined a canonical equation of state for the Taub-Matthews equation of state (i.e. $h(S,P)$
where $S$ is entropy), we have not defined a relationship with temperature $T$,
and we cannot compute specific heat capacities and subsequently $\Gamma$. Hence
the need for the proxy $\Gamma_{\text{eq}}$.} \citep[see,
e.g.][]{mignoneEquationStateRelativistic2007}:
\begin{equation}
  \Gamma_{\text{eq}} = \frac{ h - c^2}{h - c^2 - \Theta},
\end{equation}
This relationship, along with the enthalpy and speed of sound, for ideal gases
with $\Gamma=4/3$ and $\gamma=5/3$, the Synge gas, and the Taub-Matthews
equation of state is shown in Fig.~\ref{fig:eos_comparison}.

\subsection{Spatial and Temporal Discretizations}\label{sec:srhd_in_dg}
\label{sec:discontinuous_galerkin_methods}

In this work, spatial discretization of the hyperbolic conservation law,
Eq.~\ref{eq:srhd_conservation_laws}, is performed using a
discontinuous-Galerkin method in a similar fashion as was proposed
by~\cite{nunez-delarosaHybridDGFV2018}, following on the influential
sequence~\cite{cockburn1989-2,cockburn1989-3,cockburn1990-4,cockburn1998-5}.
The discontinuous-Galerkin method requires a mesh defined as the subdivision of
the domain into non-overlapping hexahedral ($3D$) or quadrilateral ($2D$) cells
denoted $\Omega_k \subset \Omega \subset \mathbb{R}^d$.  The approximation of
the conserved variables on cell $\Omega_k$ is written
\begin{equation} \label{eqn:interpolation}
\mathbf{U}(\mathbf{x}) \approx \mathbf{U}^h(\mathbf{x}) = \sum_{i=1} \mathbf{U}_i \phi_i(\mathbf{x}) \quad \mathbf{x} \in \Omega_k 
\end{equation}
where the set $\{\phi_i(\mathbf{x})\}$ is a linearly independent basis that
spans a polynomial space of fixed order on element $\Omega_k$. Lagrange polynomials are employed here, where the nodal points are denoted as $\mathbf{x}_j$ such that 
\begin{equation}
\phi_i(\mathbf{x}_j) = \delta_{ij}
\end{equation}
where $\delta$ is the Kronecker delta function.
Globally, $\mathbf{U}^h$ is defined as a piecewise
polynomial function with discontinuities permitted at cell boundaries. The restriction of the numerical solution to a cell $\Omega_k$ is denoted $\mathbf{U}_k^h$.

On each cell the approximate solution to Eq.~\ref{eq:srhd_conservation_laws} is
computed by enforcing that the residual is orthogonal to the test space,
defined in the Galerkin fashion. Practically, after integration by parts, this
implies the satisfaction of the weak form 
\begin{equation}
\label{eqn:discrete_hyperbolic_conservation_law}
\int_{\Omega_k} \frac{\partial \mathbf{U}^h}{\partial t} \phi(\mathbf{x}) d \mathbf{x} + \oint_{\partial \Omega_k} \overline{\mathcal{F}[\mathbf{W}^h(\mathbf{U})] \cdot \mathbf{n}} \phi (\mathbf{x}) ds - \int_{\Omega_k} \mathcal{F}[\mathbf{W}^h(\mathbf{U})] \cdot \nabla \phi(\mathbf{x}) d \mathbf{x} = 0, \quad \forall \phi \in \{\phi_i\} 
\end{equation}
on each cell.  The second term is the integral of normal flux over the surface
of an element. The solution at cell interfaces is double-valued as indicated by
the overline; one value corresponding to the data inside the cell, the other
from the neighboring cell. As such, the solution is discontinuous and the flux
must be computed using a Riemann solver in a fashion similar to the finite
volume method. We have implemented two \emph{approximate} Riemann solvers: HLL
and HLLC, discussed in \S\ref{sec:riemann_solvers}. Beyond the choice of
Riemann solver, the discrete conservation law,
Eq.~\ref{eqn:discrete_hyperbolic_conservation_law} can admit a range of
different basis orders. A first order basis (e.g piecewise constant) will
eliminate the contribution of $\int_{\Omega_h}
\mathcal{F}[\mathbf{W}(\mathbf{U})] \cdot \nabla \phi(\mathbf{x}) d
\mathbf{x}$, resulting in a scheme equivalent to a first order finite volume
discretization. Moving to higher order bases (e.g. piecewise linear, etc.) will
introduce the need to provide additional stabilization (e.g. dissipation) at
discontinuities and shocks.  For this we use the  Moe limiter from
\citet{moeSimpleEffectiveHighOrder2015} and the minmod limiter
\citep{vanleerUltimateConservativeDifference1979} as well as the physicality
enforcing operator tailored for relativistic hydrodynamics that we discuss in
detail in \S\ref{sec:physicality_enforcing_operator}.

Before the integrals in Eq.~\ref{eqn:discrete_hyperbolic_conservation_law} can be
computed, the primitive variables must be calculated for use in the numerical
flux. There are different options for computation: interpolate conserved and
compute primitives at quadrature points, versus compute primitives at nodal
points and interpolate. In Newtonian hydrodynamics, the primitive variables,
$\mathbf{W}$, can be recovered algebraically from the conserved state. As such,
it is straightforward to interpolate the conserved quantities to the required
quadrature point and recover the necessary primitive quantities to construct
the flux. In \emph{relativistic} hydrodynamics, such an algebraic recovery of
the primitive quantities does not exist; prior work \citep[see
e.g.][]{beckwithSecondorderGodunovMethod2011} has demonstrated that, in the
context of finite volume schemes, it is necessary to interpolate
\emph{primitive} variables (rather than conserved quantities) in order to
ensure that the state remains physical (e.g. $|\mathbf{v}|^2 < c^2$, $\rho >
0$, $P > 0$). Here, we follow a similar procedure: the primitive state is
computed from the conserved state at the basis points and then interpolated to
quadrature points in order to compute fluxes. In addition to enhanced
stability, this minimizes the number of calls to the method that
recovers the primitive variables from the conserved state, minimizing the
impact that this routine has on overall algorithm performance (see
\S\ref{sec:conserved_to_primitive} for further discussion). Thus, the first
step in the assembly is to compute the primitives at nodal points:
\begin{equation}
\mathbf{W}_i = p(\mathbf{U}_i)
\end{equation}
where $p$ computes the primitive variables from the conserved (see
Sec.~\ref{sec:conserved_to_primitive} for specific details). With this
expression, the primitives are easily interpolated to points within the cell
using Eq.~\ref{eqn:interpolation}, yielding the primitive approximation
$\mathbf{W}^h(\mathbf{x}) = \sum_{i} \mathbf{W_i}\phi_i(\mathbf{x})$. Thus a
nonlinear conserved-to-primitive solve is required at each nodal point. 

The numerical quadrature for the volumetric contributions of the fluxes are computed as
\begin{align}
\int_{\Omega_k}\mathcal{F}[\mathbf{W}^h(\mathbf{U})] \cdot \nabla \phi(\mathbf{x}) d \mathbf{x} 
& \approx 
\sum_{q} w_q \mathcal{F}[\mathbf{W}^h(\mathbf{x}_q)] \cdot \nabla \phi(\mathbf{x}_q) \label{eqn:quad-vol-flux}
\end{align}
and the surface fluxes on the interface shared by $\Omega_k$ and $\Omega_{k'}$ are
\begin{align}
\int_{\partial \Omega_k \cap \partial\Omega k'}  \overline{\mathcal{F}[\mathbf{W}^h(\mathbf{U})] \cdot \mathbf{n}} \phi (\mathbf{x}) ds  
& \approx
\sum_q \omega_q  \overline{\mathcal{F}(\mathbf{W}_k^h(\mathbf{x}_q),\mathbf{W}_{k'}^h(\mathbf{x}_q)) \cdot \mathbf{n}} \phi (\mathbf{x}_q). \label{eqn:eqn:quad-surface-flux}
\end{align}
Here it is understood that the quadrature rules are defined with respect to the
domain of integration. The volumetric term (Eq.~\ref{eqn:quad-vol-flux})
requires evaluation of the flux at each quadrature point while the surface term
(Eq.~\ref{eqn:eqn:quad-surface-flux}) requires evaluation of the numerical flux
from cell $k$ and the neighbor $k'$ at each quadrature point.

The temporal discretization we employ uses a multi-stage strong-stability
preserving (SSP) Runge-Kutta time integrator similar to that described
in~\cite{cockburn1989-2,cockburn1989-3,cockburn1990-4,cockburn1998-5}. SSP time discretization methods were designed to ensure
nonlinear stability properties in the numerical solution of spatially
discretized hyperbolic partial differential equations, such as
Eq.~\ref{eqn:discrete_hyperbolic_conservation_law}. These methods assume that
there is a time-step, $\Delta t_{FE}$ such that forward-Euler condition:
\begin{equation}
\label{eqn:forward_euler_condition}
|| \mathbf{U} + \Delta t \mathcal{F}[\mathbf{W}(\mathbf{U})] || \le || \mathbf{U} ||\;\; \mathrm{for}\;\; 0 \le \Delta t \le \Delta t_{FE}
\end{equation}
is satisfied for all $\mathbf{U}$. An explicit Runge-Kutta (ERK) method is
called SSP if the methods can be rewritten
as a convex combination of forward Euler methods and the estimate $||
\mathbf{U}^{n+1} || < || \mathbf{U}^n ||$ holds for the numerical solution of
Eq.~\ref{eqn:discrete_hyperbolic_conservation_law} whenever the condition given
in Eq.~\ref{eqn:forward_euler_condition} holds and $\Delta t \le
\mathcal{C}_{SSP} \Delta t_{FE}$, where $\mathcal{C}_{SSP}$ is known as the
SSP-coefficient. The convex combination above ensures that the
strong stability property is also satisfied by the intermediate stages in a
Runge-Kutta method \citep[
see][]{gottliebStrongStabilityPreserving2011,gottliebStrongStabilityPreserving2015}.
This may be desirable in many applications, notably in simulations that require
positivity
\citep{ferracinaExtensionAnalysisShuOsher2005,ferracinaStepsizeRestrictionsTotalVariationDiminishing2004,higuerasStrongStabilityPreserving2004,higuerasRepresentationsRungeKutta2005}.
In this work, we make use of the second and third order schemes found in
\citet{shuEfficientImplementationEssentially1989}, which were proved to be
optimal in \citet{gottliebTotalVariationDiminishing1998}.

\subsection{Computation of the Surface Flux}\label{sec:riemann_solvers}
The surface flux contributions on the interface shared by $\Omega_k$ and
$\Omega_{k'}$ require the evaluation of (Eq.~\ref{eqn:eqn:quad-surface-flux}):
\begin{align}
\sum_q \omega_q  \overline{\mathcal{F}(\mathbf{W}_k^h(\mathbf{x}_q),\mathbf{W}_{k'}^h(\mathbf{x}_q)) \cdot \mathbf{n}} \phi (\mathbf{x}_q)
\end{align}
In the method presented here, this is accomplished by use of an approximate
Riemann solver, of which we have implemented the relativistic HLL and HLLC variants due to
\citet{schneiderNewAlgorithmsUltrarelativistic1993} and
\citet{mignoneHLLCRiemannSolver2005}. Both of these approximate Riemann solvers
require an estimate of the maximum and minimum wavespeeds on either side of
the interface, which we compute through the maximum and minimum eigenvalues of
$\partial \mathbf{F}/\partial \mathbf{U}$ \citep{mignoneHLLCRiemannSolver2005}:
\begin{equation}
  \lambda_\pm (\mathbf{W} ) = \frac{ v_x \pm \sqrt{ \sigma_s \left ( c^2 - v^2_x + c^2 \sigma_s \right ) } }
  {1 + \sigma_s}
\end{equation}
where 
\begin{equation}
  \sigma_s = c_s^2/\left [ \gamma^2 \left ( c^2 - c_s^2 \right ) \right ].
\end{equation}
We compute $\lambda_\pm (\mathbf{W} )$ for every $\mathbf{W}_k^h(\mathbf{x}_q)$
and $\mathbf{W}_{k'}^h(\mathbf{x}_q))$ to find the maximum and minimum
wavespeeds at each surface quadrature point across interface:
\begin{align}
  \lambda_L & = \min{ \left ( \lambda_{-} \left ( \mathbf{W}_{k}^h(\mathbf{x}_q) \right ) ,  
                              \lambda_{-} \left ( \mathbf{W}_{k'}^h(\mathbf{x}_q) \right ) \right ) } \\
  \lambda_R & = \max{ \left ( \lambda_{+} \left ( \mathbf{W}_{k}^h(\mathbf{x}_q)\right ) ,  
                              \lambda_{+} \left ( \mathbf{W}_{k'}^h(\mathbf{x}_q) \right ) \right ) }.
\end{align}

\subsection{Physicality Enforcing Operator}\label{sec:physicality_enforcing_operator}

While using $0^{\text{th}}$ order polynomials for a relativistic hydrodynamics
discontinuous-Galerkin method is guaranteed to produce a physical conserved
state after every flux update even with shocks when using a
local-extremum-diminishing numerical fluxes such as HLL, higher order bases can
introduce spurious oscillations and non-physical conserved states within cells
around shocks (see
\cite{wuPHYSICALCONSTRAINTPRESERVINGCENTRALDISCONTINUOUS2016}). To resolve this
issue, an operator is needed to smooth the solution within a cell. Taking
inspiration from the limiter presented in
\cite{moeSimpleEffectiveHighOrder2015}, we present here a smoothing procedure
that enforces physical conserved states within a cell with a physical volume
average.

Following \cite{riccardiPrimitiveVariableRecovering2008} and
\cite{wuPHYSICALCONSTRAINTPRESERVINGCENTRALDISCONTINUOUS2016}, a conserved
state that satisfies
\begin{equation}
  \label{eq:physicality}
  D > 0, \quad q\left(\mathbf{U}\right) \equiv E/c^2 - \sqrt{ D^2 - |\mathbf{M}/c|^2} > 0,
\end{equation}
is a physically admissible state as long as the specific energy $e(\rho ,p)$ is continuously
differentiable under the chosen equation of state. If a conserved state
satisfies Eq.~\ref{eq:physicality}, the state can be inverted for a primitive
state with positive density and pressure  with a velocity less than $c$. Since
the space of permissible conserved states under Eq.~\ref{eq:physicality} is
convex (i.e. any conserved state interpolated between two physically
permissible conserved states is also physically permissible
\citep{wuPHYSICALCONSTRAINTPRESERVINGCENTRALDISCONTINUOUS2016}), we can use the
same strategies from \cite{moeSimpleEffectiveHighOrder2015} in a simple
smoothing procedure to enforce physicality within a discontinuous-Galerkin cell.
From a high level, we apply an operator to average nodal points within a cell
towards a physical volume average.

Before enforcing physicality within cells, we first screen for cells with non-physical nodal points
by checking that all conserved states at the nodal points -- $\mathbf{U}_i$ --
satisfy Eq.~\ref{eq:physicality}. If any point fails, we flag the cell as
needing smoothing to ensure that all points are physical. We then check that the
cell volume average $\overline{\mathbf{U}}$ of the conserved state satisfies
Eq.~\ref{eq:physicality}. As long as the cell volume average is physical,
a smoothing factor can be found that ensures physicality without changing the global
conserved quantities. If the cell volume average is not physical, then the
nodal points cannot be made physical through the physicality-enforcing operator without
changing the volume average.

To enforce physicality within a cell, we first seek a smoothing factor $s \in [0,1]$ such that
the smoothed states
\begin{equation}
  \label{eq:smoothed_states}
  \tilde{\mathbf{U}}_i = s \mathbf{U}_i + \left ( 1 - s \right ) \overline{\mathbf{U}}
\end{equation}
at all nodal points in the cell satisfy Eq.~\ref{eq:physicality}. At each
point in the cell, we find the largest smoothing factor such that
\begin{equation}
  \label{eq:smoothed_state_physicality}
  \tilde{D}_i > 0  \quad \tilde{q}_i \equiv \tilde{E}_i/c^2 - \sqrt{ \tilde{D}_i^2 + \left (|\tilde{\mathbf{M}}_i|/c \right )^2} > 0.
\end{equation}
If we assume that $\overline{\mathbf{U}}$ is physical, then $s :=0$ would
lead to a physical $\tilde{\mathbf{U}}$, so we can assume that such a
smoothing factor $s_i \geq 0$ exists. We find this factor in two stages. 

In the first stage, we compute an intermediate stage smoothing factor $s_i^{(1)}$
for each nodal point that ensures a positive $D$ and $E$. We solve
\begin{align}
  \label{eq:first_stage_positivity}
  \tilde{D}^{(1)}_i &= s^{(1)}_{i,D} D_i + \left ( 1 - s^{(1)}_{i,D} \right ) \overline{D} > 0 \\
  \tilde{E}^{(1)}_i &= s^{(1)}_{i,E} E_i + \left ( 1 - s^{(1)}_{i,E} \right ) \overline{E} > 0
\end{align} for the largest $s^{(1)}_{i,D},s^{(1)}_{i,E}\in [0,1]$
that satisfies the constraints and compute an intermediate smoothing factor
$s^{(1)}_i = \min { \left ( s^{(1)}_{i,D},s^{(1)}_{i,E} \right ) }$. We use
$s^{(1)}_i$ to compute an intermediate smoothed state 
\begin{equation}
  \label{eq:first_stage_smoothed_state}
\tilde{\mathbf{U}}^{(1)}_i = s^{(1)}_i \mathbf{U}_i + \left ( 1 - s^{(1)}_i \right ) \overline{\mathbf{U}}
\end{equation}
so that we ensure that $\tilde{D}$ and $\tilde{E}$ are positive.

In the second stage, we compute a second stage smoothing factor $s^{(2)}_{i}
\in [0,1]$ such that
\begin{equation}
  \label{eq:second_stage_q}
  \tilde{q}^{(2)}_i = \tilde{E}^{(2)}_i/c^2 - \sqrt{ \left( \tilde{D}^{(2)}_i \right )^2 + \left( |\tilde{\mathbf{M}}^{(2)}_i|/c \right )^2} > 0.
\end{equation}
where $\tilde{\mathbf{U}}^{(2)}_i = s^{(2)}_i \mathbf{U}^{(1)}_i + \left
( 1 - s^{(2)}_i \right ) \overline{\mathbf{U}}$ is the second smoothed
state. Note that since $s^{(2)}:=0$ leads to
$\tilde{\mathbf{U}}^{(2)}:=\overline{\mathbf{U}}$, we know that an acceptable
smoothing factor exists. 
Solving Eq.~\ref{eq:second_stage_q} can be simplified by noting that $\tilde{E}^{(2)}$ is
positive for any choice of $s^{(2)}_i \in [0,1]$ since $\tilde{E}^{(1)}$
and $\overline{E}$ are both positive (for the same reasons, $\tilde{D}^{(2)}$ is also always positive).
We can rewrite Eq.~\ref{eq:second_stage_q} as 
\begin{gather}
  \label{eq:second_stage_quatratic}
  \left (\tilde{E}^{(2)}_i/c^2\right )^2 > \left( \tilde{D}^{(2)}_i \right )^2 + \left( |\tilde{\mathbf{M}}^{(2)}_i|/c \right )^2   \\
  a \left ( s^{(2)}_i \right )^2 + b s^{(2)}_i + c > 0
\end{gather}
where
\begin{align}
  \label{eq:second_stage_coefficients}
  a & = \frac{1}{c^4}\left (\tilde{E}^{(1)}_i - \overline{E} \right )^2
                   - \left (\tilde{D}^{(1)}_i - \overline{D} \right )^2
      -\frac{1}{c^2} \left |\tilde{\mathbf{M}}^{(1)}_i - \overline{\mathbf{M}}\right |^2 \\
  b & = \frac{2}{c^4} \overline{E} \left (\tilde{E}^{(1)}_i - \overline{E} \right )
                  - 2 \overline{D} \left (\tilde{D}^{(1)}_i - \overline{D} \right )
      -\frac{2}{c^2} \overline{\mathbf{M}} \cdot \left (\tilde{\mathbf{M}}^{(1)}_i - \overline{\mathbf{M}}\right )^2 \\
  c & = \frac{1}{c^4} \overline{E}^2
                    - \overline{D}^2
      -\frac{1}{c^2} \left |\overline{\mathbf{M}}\right |^2.
\end{align}
Since $\overline{\mathbf{U}}$ is physical, $s^{(2)}_i:=0$ must satisfy the
inequality. Note that the quadratic can only have at most one root within
[0,1]; if it had two roots, then either $s^{(2)}_i:=0$ and $s^{(2)}$
do not satisfy the inequality, implying that $\overline{\mathbf{U}}$ is
unphysical, or that both satisfy the inequality and that some interior
$s^{(2)}_i \in [0,1]$ do not satisfy the inequality, implying that the
space of physical conserved states is not convex, both of which are
contradictions.  If there are no roots within $[0,1]$, since $s^{(2)}_i:=0$
satisfies the inequality, $s^{(2)}_i:=1$ must as well, so $1$ would be
the largest acceptable second stage smoothing factor.

In the case that there is just one root, then since $s^{(2)}_i:=0$
satisfies the inequality, the coefficient $a$ must be negative or $0$ (which is
the simple linear case), and only the root
\begin{equation}
  \label{eq:second_stage_solution}
  s^{(2)}_i = \frac{ -b - \sqrt{b^2 - 4 a c}}{2a}
\end{equation}
can fall within $[0,1]$, and so we only need to compute this root to find the
largest smoothing factor for this nodal point. The final smoothing factor for
this nodal point is $s_i = s^{(1)}_i s^{(2)}_i$, which ensures that any $s \leq
s_i$ chosen will satisfy Eq.~\ref{eq:smoothed_state_physicality}.  After
computing $s_i$ for each nodal point in the cell, we compute the final
smoothing factor for the cell using $s = \min s_i$, which we use to compute
$\tilde{\mathbf{u}}$ using Eq.~\ref{eq:smoothed_states}.

The procedure for our physicality-enforcing operator goes as follows
\begin{enumerate}
  \item We flag cells with nodal points with conserved states that violate
    Eq.~\ref{eq:physicality} as cells with non-physical nodal points.
  \item We check that the volume average within a flagged cell satisfies equation
    \ref{eq:physicality}, which guarantees that the smoothing procedure will
    enforce physicality within the cell.
  \item For each point in a flagged cell, we compute the largest smoothing factor
    $s_i$ that will guarantee that the new smoothed state will satisfy
    Eq.~\ref{eq:smoothed_state_physicality}. For each nodal point, the procedure goes as:
    \begin{enumerate}
      \item We compute the first stage smoothing factor 
        $s^{(1)}_{i,D}$ and $s^{(1)}_{i,E}$ to ensure positivity of $D$ and $E$ by solving for them in
        Eq.~\ref{eq:first_stage_positivity}.
      \item We compute the first stage smoothing factor $s^{(1)}_i = \min {
          s^{(1)}_{i,D},s^{(1)}_{i,E} }$ and use this to compute
        the intermediate smoothed state $\tilde{\mathbf{U}}^{(1)}$ using
        Eq.~\ref{eq:first_stage_smoothed_state}.
      \item We then check whether $\tilde{\mathbf{U}}^{(1)}$ satisfies equation
        \ref{eq:physicality}, in which case we use $s_i =
        s^{(1)}_i$.
      \item If not, we compute $s^{(2)}_i$ by solving the quadratic
        described in Eq.~\ref{eq:second_stage_quatratic} and
        Eq.~\ref{eq:second_stage_coefficients} using the root for $s^{(2)}_I$
        in Eq.~\ref{eq:second_stage_solution}. The smoothing factor for this nodal
        point is then $s_i = s^{(1)}_i s^{(2)}_i$.
    \end{enumerate}
  \item  We compute a final smoothing factor for each cell using  $s = \min
    s_i$, which allows us to compute the smoothed state $\mathbf{U}_i$ at
    each nodal point using Eq.~\ref{eq:smoothed_states}.
\end{enumerate}
As long as the volume average conserved state $\overline{\mathbf{U}}$ is
physical, this procedure will produce the physical conserved state
$\tilde{\mathbf{U}}_i$.


\section{Recovery of Primitive Variables}\label{sec:conserved_to_primitive}

Although the conservation laws in relativistic hydrodynamics are similar to
those in Newtonian hydrodynamics, the inclusion of the Lorentz factor in
conservation of mass, momentum, and energy adds complexity to the equation set
in several ways that complicate recovery of primitive variables from conserved variables. Primarily, the Lorentz factor couples every
conserved variable with the velocity in all directions. While adding a
transverse velocity to a non-relativistic flow will not affect  longitudinal
evolution, in demonstration of Galilean invariance, a transverse velocity in a
relativistic flow contributes to the apparent density, momentum, and energy,
fundamentally modifying the dynamics. Additionally, the inclusion of the
Lorentz factor leads to a non-linear relationship between the primitive and
conserved variables. For even simple choices of equation of state, recovering
the primitive state from the conserved state (i.e. inverting
Eq.~\ref{eq:primitive_to_conserved})
requires finding the roots of cubic or higher order
polynomials. Last, the relativistic hydrodynamics equations (and causality)
require the three-velocity to be bounded by the speed of light, with superluminal velocities
leading to complex Lorentz factors.  For highly relativistic flows close to the
speed of light, we are often limited by machine precision when representing
small changes in the three-velocity that equate to large changes in the Lorentz
factor. For these reasons, the stability and fidelity of any scheme for
relativistic hydrodynamics is fundamentally tied to that of the scheme used to
compute primitive variables from conserved quantities. As a result, a wide
variety of schemes, including but not limited to those presented in
\citet{schneiderNewAlgorithmsUltrarelativistic1993,ryuEquationStateNumerical2006,riccardiPrimitiveVariableRecovering2008},
have been described in the literature. Each of these options has
its advantages and disadvantages from a physical fidelity,
stability, and robustness standpoint; however, as far as we are aware, the
performance of these different formulations has not previously been examined
from a performance portability perspective, as we do here.

We consider two different approaches to recovering the primitive variables from
conserved quantities: an analytical approach and an iterative approach. We then develop
both of these methods for the ideal gas and Taub-Matthews equations of state to
give four algorithms in all. In formulating these, we use the dimensionless
variables
\begin{equation}
  \xi = \frac{M}{D c} \quad \text{and} \quad \eta = \frac{E}{D c^2}.
\end{equation}
This rescaling aids with reducing issues due to large differences in numbers,
although this does not eliminate issues of near-speed-of-light velocities.

\subsection{Ideal Gas Equation of State}\label{sec:ideal_gas_primitive_solver}

In the case of the ideal gas equations of state, the primitive variables can be
recovered from the conserved quantities by solving the roots of a quartic
equation. One approach demonstrated by \citet{ryuEquationStateNumerical2006}
computes the analytic solution to a quartic polynominal in $\beta = v/c$. For
completeness, we restate this method here in terms of the dimensionless
parameters $\xi$ and $\eta$, which allows us to keep $c$ throughout the set of
equations. 

As shown in \citet{schneiderNewAlgorithmsUltrarelativistic1993}, the solution for the special
relativistic velocity $\beta$ can be found from the roots of the quartic
polynomial
\begin{equation}
  a_3 \beta^4 + a_2 \beta^2 + a_1 \beta + a_0 = 0
\end{equation}
where the coefficients are given by
\begin{align}
  a_3 &= \frac{ -2 \Gamma ( \Gamma -1 ) \xi \eta }{ (\Gamma -1)^2 ( \xi^2 + 1) }\\
  a_2 &= \frac{ \Gamma^2 \eta^2 + 2 (\Gamma -1 ) \xi^2 - (\Gamma -1 )^2 }{ (\Gamma -1)^2 ( \xi^2 + 1) }\\
  a_1 &= \frac{ -2 \Gamma \xi \eta }{ (\Gamma -1)^2 ( \xi^2 + 1) }\\
  a_0 &= \frac{ \xi^2 }{ (\Gamma -1)^2 ( \xi^2 + 1) }.
\end{align}
Only one root of the polynomial provides a physical $\beta \in [0,1)$. The root
can be found using a root-finding method or analytically \citep{ryuEquationStateNumerical2006} through: 
\begin{equation}
  \beta = \frac{-B + \sqrt{B^2 - 4 C} }{2}
\end{equation}
where
\begin{align}
  B & =\frac{1}{2} \left ( a_3 + \sqrt{ a_3^2 - 4 a_2 + 4 x} \right ) \\
  C & = \frac{1}{2} \left ( x - \sqrt{ x^2 - 4 a_0 } \right )
\end{align}
We then have that:
\begin{equation}
x =
  \begin{cases}
    2 \left ( R^2 + T \right)^{2/3} \cos{ \left [ \frac{1}{3} \tan^{-1}{ \left (  \frac{\sqrt{-T}}{R} \right )} \right ]} - i_1/3 & \text{ if } T <0 \\ 
     \left (R + \sqrt{T} \right )^{1/3} + \left ( R - \sqrt{T} \right )^{1/3} - i_1/3 & \text{ otherwise }
  \end{cases}
\end{equation}
where $R$, $S$, and $T$ are found from
\begin{align}
  R & = \frac{1}{54} \left ( 9 i_2 i_2 - 27 i_3 - 2 i_1^3 \right ) \\
  S & = \frac{1}{9} \left (3 i_2 - a_2^2 \right ) \\
  T & = R^2 + S^3
\end{align}
where 
\begin{align}
  i_1 & = -a_2 \\
  i_2 & = a_3  a_1 - 4a_0 \\
  i_3 & = 4 a_2 a_0 - a_1^2 - a_3^2 a_0. \\
\end{align}
With a solution for $\beta$, the rest of the primitive variables can be recovered using
\begin{align}
  \rho & = D \sqrt{1 - \beta^2} \\
  \mathbf{v} & = \frac{\beta}{\xi D} \mathbf{M} \\
  P & = (\Gamma -1 )\left( E - \mathbf{M} \cdot \mathbf{v} - \rho c^2 \right).
\end{align}

An alternative strategy for recovering the primitive variables from conserved
quantities is to utilize an iterative solver to find the roots.  Exploring the
iterative approach, we used an iterative solver following the recovery method
presented in \citet{riccardiPrimitiveVariableRecovering2008}. This solver has
two main advantages. First it uses a proxy for the velocity that scales more
evenly from weakly to highly relativistic flows. Second, the resulting quartic
polynomial can be solved using the Newton-Raphson method, which it typically
more robust, accurate, and faster even using several iterations due to avoiding
the slow and imprecise square roots and inverse tangents in the analytic
solver.

Instead of recovering the primitives by solving for velocity, Lorentz factor,
or pressure, we instead solve for a proxy of the velocity,  $w$, where 
\begin{equation}
  \label{eq:iterative_quartic_unknown}
  u = \frac{ 2 w}{1 + w^2}.
\end{equation}
We solve for $w \in (0,1)$ by finding the root within $(0,1)$ of the quartic
polynomial 
\begin{equation}
  \label{eq:iterative_quartic}
  P(w) = 
      \left ( \alpha - 1      \right ) \xi w^4 
  - 2 \left ( \alpha \eta + 1 \right ) w^3 
  + 2 \left ( \alpha + 1      \right ) \xi w^2 
  - 2 \left ( \alpha \eta - 1 \right ) w 
  +   \left ( \alpha - 1      \right ) \xi,
\end{equation}
where $\alpha = \Gamma/(\Gamma -1 )$. Within the range $w \in (0,1)$, the
equation $P(w)=0$ has only one root. While $P(w)=0$ could be solved
analytically using the same method for our analytical solver, the
Newton-Raphson method is simpler and often quicker, since it only requires
addition and multiplication and coefficients of the polynomial can be reused
across iterations. We also find that the Newton-Raphson method always
converges to the root in $(0,1)$ as long as the initial guess is in $(0,1)$,
which is consistent with \citet{riccardiPrimitiveVariableRecovering2008}.  This
obviates the need for a bounded root solver. For reasonably relativistic flows
with $\gamma < 10$, this may only take 5 iterations to recover $w$ to within
double floating point machine precision ($\Delta w \sim 10^{-16}$).

When $\xi$ is very small, a cubic approximation for a solution for $w$ can be used
\begin{equation}
  \label{eq:iterative_cubic_approximation}
  w = \frac{ \alpha -1 }{2 \left ( \alpha \eta - 1 \right ) } \xi
    + \frac{ \left (\alpha -1\right )^2 }{8 \left ( \alpha \eta - 1 \right )^4 } \left [
      \left ( \alpha +3 \right ) \left ( \alpha \eta + 1 \right ) - 4 \left ( \alpha + 1 \right ) \right ]
    \xi^3 + O(\xi^5).
\end{equation}

Generally, the  iterative solver for the ideal equation of state is more
accurate than the analytical solver. Often, the iterative solver is also
faster. Comparison between the solvers for the ideal equation of state and the
solvers for the Taub-Matthews equation of state are explored in section
\ref{sec:conserved_to_primitive_solver_comparisons}.

\subsection{Taub-Matthews Equation of State}\label{sec:conserved_to_primitive_synge}

For the Taub-Matthews equation of state, the primitive state can be recovered
from the conserved state by solving a cubic equation for $W=\gamma^2 -1$.
Following \citet{ryuEquationStateNumerical2006}, we solve for $W$ from 
\begin{equation}
  \label{eq:synge_cubic}
  W^3 + c_1 W^2 + c_2 W + c_3 = 0
\end{equation}
where 
\begin{align}
  c_1 = \frac{ \left ( \eta^2 + \xi^2 \right ) \left [ 4 \left ( \eta^2 + \xi^2 \right ) - \left ( \xi^2 + 1 \right ) \right ] - 14 \xi^2 \eta^2 }
             {2 \left ( \eta^2 - \xi^2 \right )^2 } \\
  c_2 = \frac{ \left [ 4 \left ( \eta^2 + \xi^2 \right ) - \left ( \xi^2 + 1 \right ) \right ]^2  - 57 \xi^2 \eta^2 }
             {16 \left ( \eta^2 - \xi^2 \right )^2 } \\
  c_3 = - \frac{ 9 \xi^2 \eta^2 }
             {16 \left ( \eta^2 - \xi^2 \right )^2 }.
\end{align}
Eq.~\ref{eq:synge_cubic} can be solved analytically and iteratively.
Analytically solving the cubic polynomial is straightforward compared to
solving the quartic polynomial for the ideal equation of state. The solution
for $W$ depends on the discriminant of the cubic equation
\begin{equation}
  d = Q^3 + R^2
\end{equation}
with 
\begin{align}
  Q = \frac{1}{9} \left ( 3 c_2 - c_1^2 \right ) \\
  R = \frac{1}{54} \left ( 9 c_1 c_2 - 27 c_3 - 2 c_1^3 \right ). \\
\end{align}
If $d<0$, then Eq.~\ref{eq:synge_cubic} has the solution
\begin{equation}
  W = 2 \sqrt{-Q} \cos \left ( \frac{ \iota}{3} \right ) - \frac{c_1}{3}
\end{equation}
with
\begin{equation}
  \iota = \cos^{-1} \left ( \frac{R}{\sqrt{-Q^3}} \right ).
\end{equation}
Otherwise if $d \geq 0$, then Eq.~\ref{eq:synge_cubic} has the solution
\begin{equation}
  W = - \frac{c_1}{3} + S + T
\end{equation}
with 
\begin{align}
  S = \left ( R + \sqrt{d} \right )^{1/3} \\
  T = \left ( R - \sqrt{d} \right )^{1/3}.
\end{align}

A root-finding method can also be used to recover $W$ from
Eq.~\ref{eq:synge_cubic}. As an alternative option to the analytic solution, we
use the bracketed root solver Brent's method
\citep{brentrichardp.AlgorithmsMinimizationDerivatives1973} to recover $W$. For
the Taub-Matthews equation of state, we use Brent's method instead of the
Newton-Raphson since Brent's method allows us to bracket the one non-negative
root. Unlike for the quartic polynomial solved for the ideal equation of state,
the Newton-Raphson method is not guaranteed to converge to the positive root
when using a positive initial guess, which leads to an incorrect and unphysical
recovered velocity. We first bracket the root $W$ with the region corresponding
to $\gamma \in [1,200]$, then iteratively expand the upper range if the root is
not found. For the tests explored here $\gamma = 200$ is a sufficiently high
upper bound that this rebracketing is not needed.

With $W$ recovered, the Lorentz factor and relativistic velocity can be recovered via
\begin{equation}
  \gamma = \sqrt{W  + 1 } \qquad \beta = \sqrt{ \frac{W}{W+1} }.
\end{equation}
The lab frame density $\rho$ and velocity $\mathbf{v}$ can be recovered via the
same method as the ideal equation of state.
The pressure with the Taub-Matthews equation of state is recovered via 
\begin{equation}
  P = \frac{ \left ( E - \mathbf{M} \cdot \mathbf{v} \right )^2 - \rho^2}
           { 3 \left ( E - \mathbf{M} \cdot \mathbf{v} \right ) }.
\end{equation}

\subsection{Conserved to Primitive Solver Comparisons}\label{sec:conserved_to_primitive_solver_comparisons}

\begin{figure}[hbt!]
  \includegraphics[width=\linewidth]{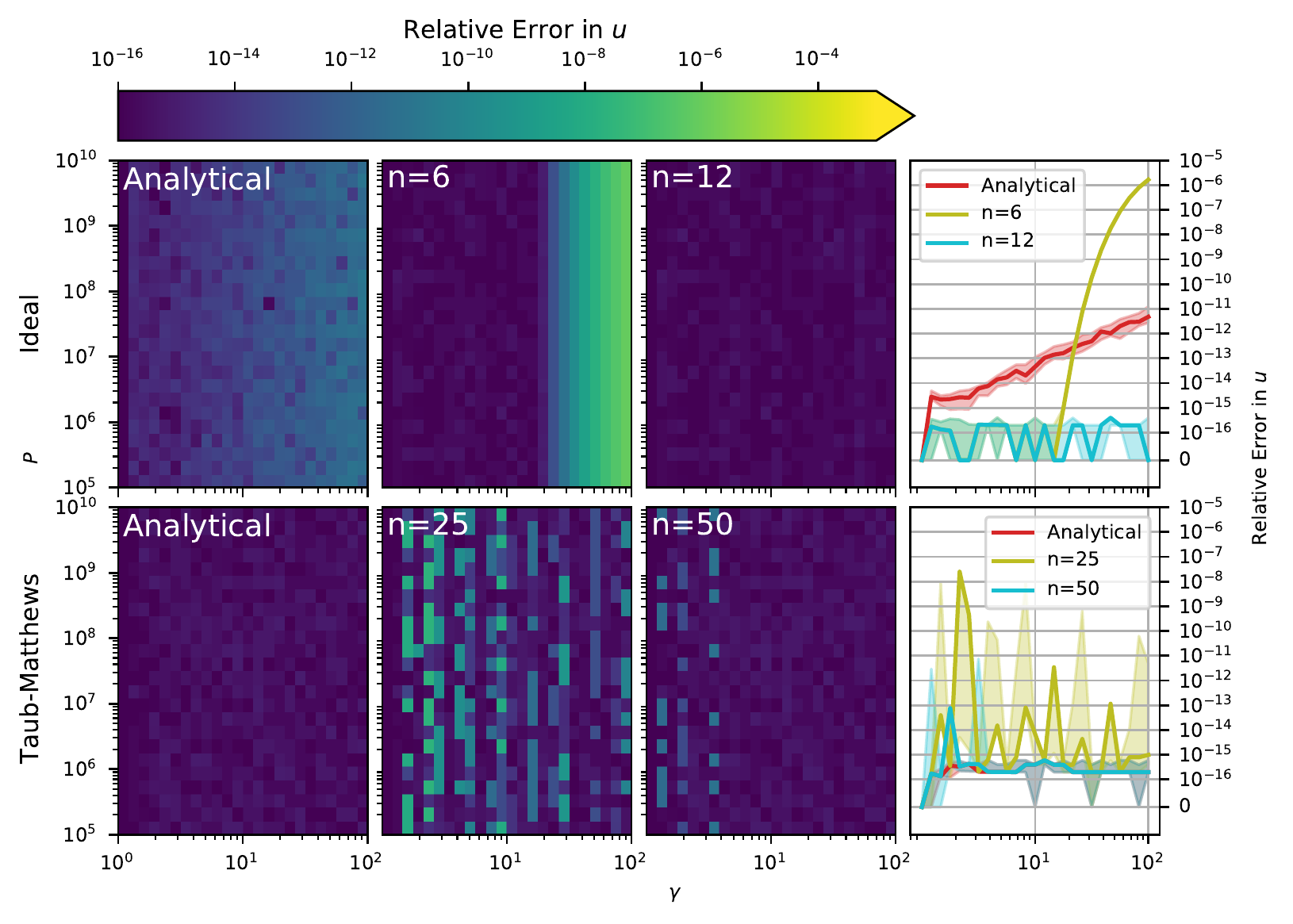}
  \caption{
    \label{fig:conserved_to_primitive_accuracy}
    Map of the error of the conserved-to-primitive solvers with the error using
    the analytical method in the left column and using varying numbers of
    iterations in the middle two columns and error of these configurations
    versus Lorentz factor in the right column. The top row shows results for
    the ideal gas, testing the iterative solver with 6 and 12 iterations, and
    the bottom row shows results for the Taub-Matthews equation of state, testing the iterative
    solver using 25 and 50 iterations.
    In all panels, $25\times25$ primitive states are tested with Lorentz
    factors varying from 1 to 100 on the $x$-axis and pressures varying from
    $10^5$ to $10^{10} \text{ N m}^{-2}$, using  $c=3\times 10^8\text{ m
    s}^{-1}$ and fixing $D=1 \text{ kg m}^{-3}$, these primitive states are
    first converted to conserved states and then converted back to a primitive
    state using the specified analytical or iterative solver. In the left three
    columns, the relative error is shown in color with the $y$-axis showing the
    pressure. In the rightmost column, the median (solid line) and first to
    third quartile (shared region) of the error sampled using different
    pressures given a specific Lorentz factor. All results in this figure are
    using the Intel compiler on CPUs.  The iterative solver for the ideal
    equation of state is more accurate than the analytic solver using just $12$
    iterations for high Lorentz factors and just $6$ iterations for low Lorentz
    factors. For the Taub-Matthews equation of state, the analytical solver is
    almost always at least or more accurate than the iterative solver.
  }
\end{figure}

Fig.~\ref{fig:conserved_to_primitive_accuracy} shows the relative error in the
recovered velocity in the ideal gas equation of state and Taub-Matthews equations of state
using the analytical method and iterative methods using varying number of
iterations. The plots are created by applying the methods on a grid of $25^2$
primitive states with $D=1 \text{ kg m}^{-3}$ and 25 logarithmically spaced pressures from $10^5$
to $10^{10} \text{ N m}^{-2}$ and 25 logarithmically spaced Lorentz factors
from $1$ to $100$, using $c=3\times 10^8 \text{m s}^{-1}$.  Each pair of
pressure and Lorentz factor is converted to a conserved state using
Eq.~\ref{eq:primitive_to_conserved} that is converted back to a primitive state
using the specified recovery method. We then compute the relative error of the
velocity in the recovered primitive state to the original velocity determined
by the Lorentz factor. 

For the ideal gas using 64 bits of floating precision, the analytical solver
recovers the velocity to $10^{-15}$ for Lorentz factors below $3$ and in some
cases recovering it exactly due to machine precision ($10^{-16}$ in this
regime). The accuracy of the analytical method decreases roughly as a power law
with increasing Lorentz factor, reaching about $10^{-10}$ at $\gamma = 100$. At
this high Lorentz factor, the relative error in recovered Lorentz factor is
$10^{-6}$, which propogates into other recovered primitives, highlighting the
need to accurate recovery of velocity for ultrarelativistic flows. In contrast,
the iterative method for the ideal gas recovers the velocity exactly or near
machine precision for Lorentz factors below $10$ in only $6$ iterations, past
which the error increases rapidly with Lorentz factor. Owing to the flexibility
of the accuracy of the iterative method, increasing the iteration count to 12
leads to recovering the velocity near machine precision for all Lorentz factors
tested. At higher Lorentz factors, the iterative solver has relatively more
difficulty in recovering the velocity due to the method recovering the velocity
from a proxy of the velocity and the slow variation of velocity at high Lorentz
factors. Small errors in the recovered velocity at high Lorentz factors amplify
to large errors in other recovered primitives. We also note that for very high
pressures at and above $10^{20} \rho c^2$, analytical method for the ideal gas
encounters imaginary numbers and fails to recover the velocity at all, whereas
the iterative solver does not fail with very high pressures.

In comparison, the cubic analytic solver for the Taub-Matthews equation of
state performs closer to machine precision across the domain of primitive
states tested. The iterative solver for the Taub-Matthews equation of state
requires many more iterations than for the ideal gas equation of state.  We
attribute this to the construction of the polynomial for the iterative solver
for the ideal equation of state, which is designed to converge in a few
iterations. The Taub-Matthews equation of state iterative solver performs worse
at lower Lorentz factors since it recovers the velocity from a proxy of the
Lorentz factor, and the Lorentz factor varies slowly at low velocities. Small
errors in the recovered Lorentz factor at sub-relativistic velocities amplify
to large errors in other recovered primitives. Generally, the iterative solver
for the Taub-Matthews equation of state is less accurate than the analytical
solver, and the high iteration counts required lead to slower performance.

\begin{figure}[hbt!]
  \centering
  \includegraphics[width=0.5\linewidth]{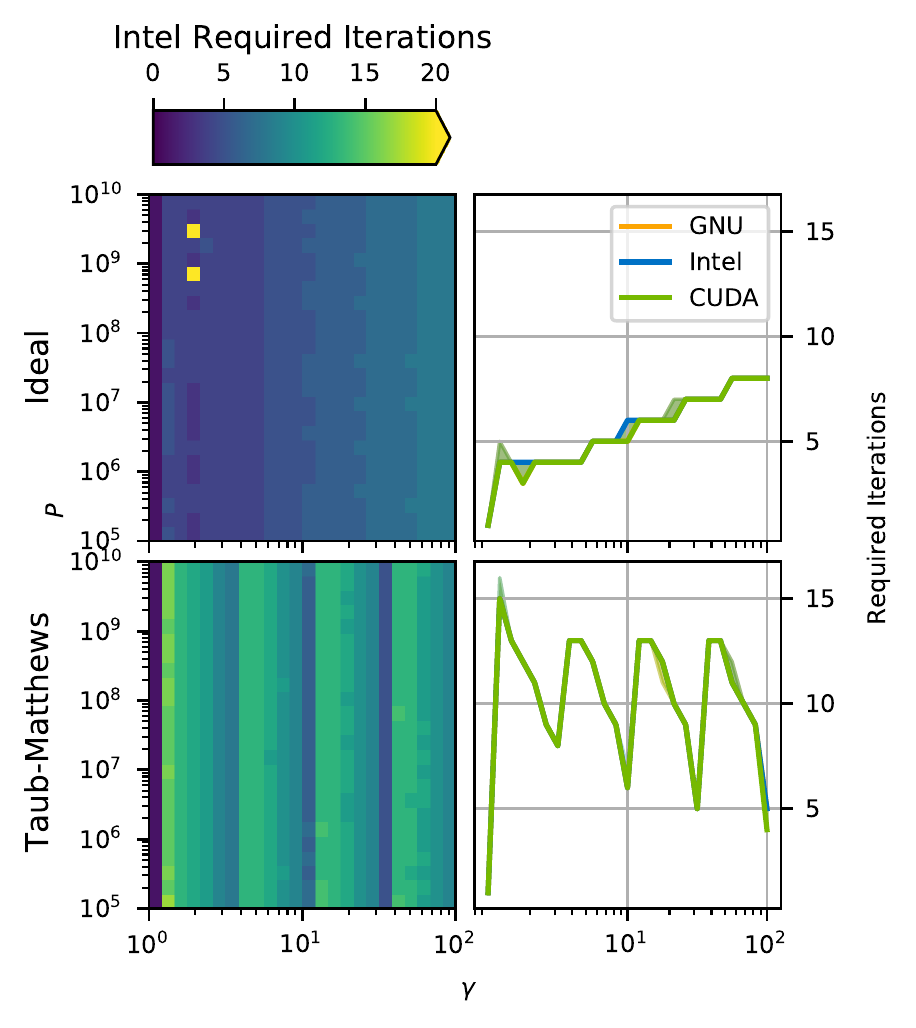}
  \caption{
    \label{fig:conserved_to_primitive_iterations} Required iterations for the
    iterative solver to reach the same accuracy as the analytical solver using
    the same primitive states as
    Fig.~\ref{fig:conserved_to_primitive_accuracy}, with results for the ideal
    gas in the top row and the Taub-Matthews equation of state in the bottom
    row. The left column shows the required iterations when compiling with the Intel compiler in
    color with Lorentz factor on the $x$ axis and pressure on the $y$ axis. For
    two primitive states the ideal analytic solver recovers the velocity
    exactly, leading the iterative solver being unable to reach the same
    accuracy, which we show in yellow.  The right column shows the median
    (solid line) and first to third quartile (shared region) of the error
    sampled using different pressures given a specific Lorentz factor, Results
    with the GNU compiler on CPUs are shown in orange, with the Intel compiler
    on CPUs with the Kokkos OpenMP backend in blue, and with the Kokkos CUDA
    backend on GPUs in green.
  }
\end{figure}

We next investigate the number of iterations required for the iterative solver
to reach accuracy parity with the analytic solver in
Fig.~\ref{fig:conserved_to_primitive_iterations}. In this figure, we test the
same grid of primitive states used in
Fig.~\ref{fig:conserved_to_primitive_accuracy}, running the iterative solver
with increasing number of iterations until it achieves greater accuracy than
the analytic solver. For some cases with the ideal gas, the analytic solver
recovers the velocity exactly, which we mark with yellow.

The number of iterations required for the iterative solvers to reach accuracy
parity depends mostly on the Lorentz factor with some variation in pressure.
The iterative solver for the ideal gas requires more iterations at higher
Lorentz factors. We attribute this to the iterative solver recovering the
primitive state by first recovering a proxy for the velocity instead of Lorentz
factor, which requires less precision to recover at low Lorentz factors.
For the primitives states tested here that the analytical solver does not
recover exactly, the ideal iterative solver requires fewer than 10 iterations
to achieve parity. We attribute the low iteration count to the one physical
root of the quartic always being the same root. 

The iterative solver required comparatively more iterations, almost always more
than 5 and upwards of 15 for low Lorentz factors. Generally more iterations are
required for lower Lorentz factors, possibly due to the solver recovering a
proxy of the Lorentz factor first, from which recovering the velocity is
sensitive to precision. The required iterations form a sawtooth with Lorentz
factors due to the physical root switching positions. 

\begin{figure}[hbt!]
  \centering
  \includegraphics[width=\linewidth]{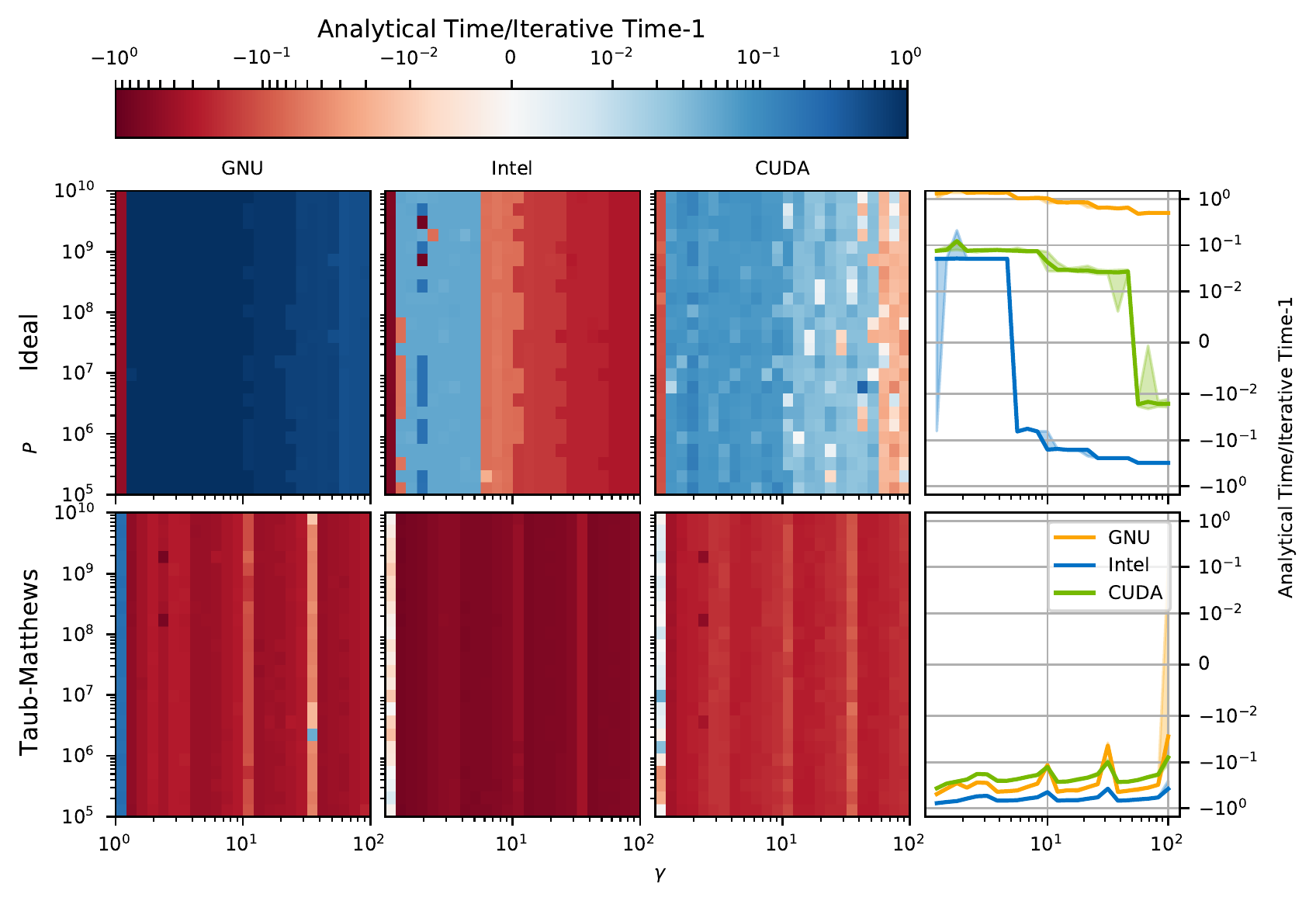}
  \caption{ \label{fig:conserved_to_primitive_timing} 
    Timing comparisons for the iterative solver to reach the same accuracy as
    the analytic solver, with comparisons as a color map in the left three
    panels and versus Lorentz factor in the rightmost panel, using the same
    primitive states as Fig.~\ref{fig:conserved_to_primitive_accuracy} with
    results for the ideal gas in the top row and the Taub-Matthews equation of state in the bottom
    row.  In all panels we compare results using the metric $\text{Analytical
    Time}/\text{Iterative Time} - 1$, where a positive value shows how much
    slower the analytical solver is as a fraction of the time the iterative
    solver takes and a negative value shows the fraction by which the
    analytical solver is faster.  The left three columns show the timing metric
    in color (blue shows where the iterative method is faster) with the Lorentz
    factor on the $x$ and the pressure on the $y$ axis, showing comparisons for
    the GNU and Intel compilers on CPUs with the Kokkos OpenMP backend and on
    GPUs with the Kokkos CUDA backend across the three columns.  The rightmost
    column shows the median (solid line) and first to third quartile (shared
    region) of the error sampled using different pressures given a specific
    Lorentz factor, showing results for all compilers tested (note that this
    does not compare timings between compilers, only the analytic against the
    iterative solver for each compiler). For the ideal equation of state,  the
    iterative solver is faster than the analytic solver under a certain
    threshold of Lorentz factor that is compiler and architecture dependent.
    The iterative solver for the Taub-Matthews equation of state is almost
    always slower than the analytic method.
  }
\end{figure}

Depending on the architecture and compiler, the iterative solver for the ideal
gas is usually faster than the analytic solver, while for the Taub-Matthews
equation of state the iterative solver is almost always slower.  We investigate
the performance of the recovery methods in
Fig.~\ref{fig:conserved_to_primitive_timing}. Using the same grid of primitive
states that we used in Fig.~\ref{fig:conserved_to_primitive_accuracy}, we
compare the run times of the analytical solvers and iterative solvers with the
number of iterations required to achieve accuracy parity,  running each of the
primitive states from Fig.~\ref{fig:conserved_to_primitive_accuracy} on $10^3$
cells with $27$ points per cell, taking an average runtime over $100$ runs
each.  We compare timings using the metric $\text{Analytical
Time}/\text{Iterative Time} - 1$, where the iterative time is with the number
of iterations required to match the analytical accuracy, in order to highlight
where the iterative solver is faster. Negative values show the fraction by
which the analytical method is faster than the iterative method while positive
values show the fraction by which the analytical solver is slower. 

For the ideal gas on CPUs using the Intel compiler, the iterative solver is
about 10\% faster than the analytical solver at Lorentz factors below $10$ and
about 10\% slower at Lorentz factors above $10$. For higher iteration counts
reaching to 10 iterations, the analytical solver begins to be faster than the
iterative solver by several percent. However, it should be noted from
Fig.~\ref{fig:conserved_to_primitive_accuracy} that in this regime the
analytical method introduces more inaccuracy to the primitive state, while the
iterative solver can recover the primitive state with much better accuracy at
the cost of performance. A red line on the right hand side shows that the
analytical solver more quickly identifies the zero velocity case, whereas the
iterative solver takes longer due the layout of the code and using the cubic
approximation from Eq.~\ref{eq:iterative_cubic_approximation} for near-zero
momenta.

Using the GNU compiler on CPUs, the iterative solver is always faster than the
analytical solver except for trivial cases. We attribute this slowdown with GNU
to the slower math functions required in the analytic solver.

For GPUs, the iterative solver for the ideal gas is faster than the analytical
solver by several percent for all but the trivial case and Lorentz factors
above $60$. This is despite the potential for the kernel to branch at every
point if different points require different numbers of iterations, although
these timing tests do not exercise this possibility. The timing disparity  may
be due to the `sqrt` operation in the analytical solver, which is more
optimized on CPUs compared to GPUs.

Considering the Taub-Matthews equation of state, the iterative solver is almost
always slower than the analytical solver. This is expected from the larger
number of iterations needed for the iterative solver to reach parity with the
analytical solver. The performance difference is largest on the Intel compiler,
where the optimized math functions allow good performance for the analytical
solver.

\begin{figure}[hbt!]
\begin{center}
  \includegraphics[width=0.5\linewidth]{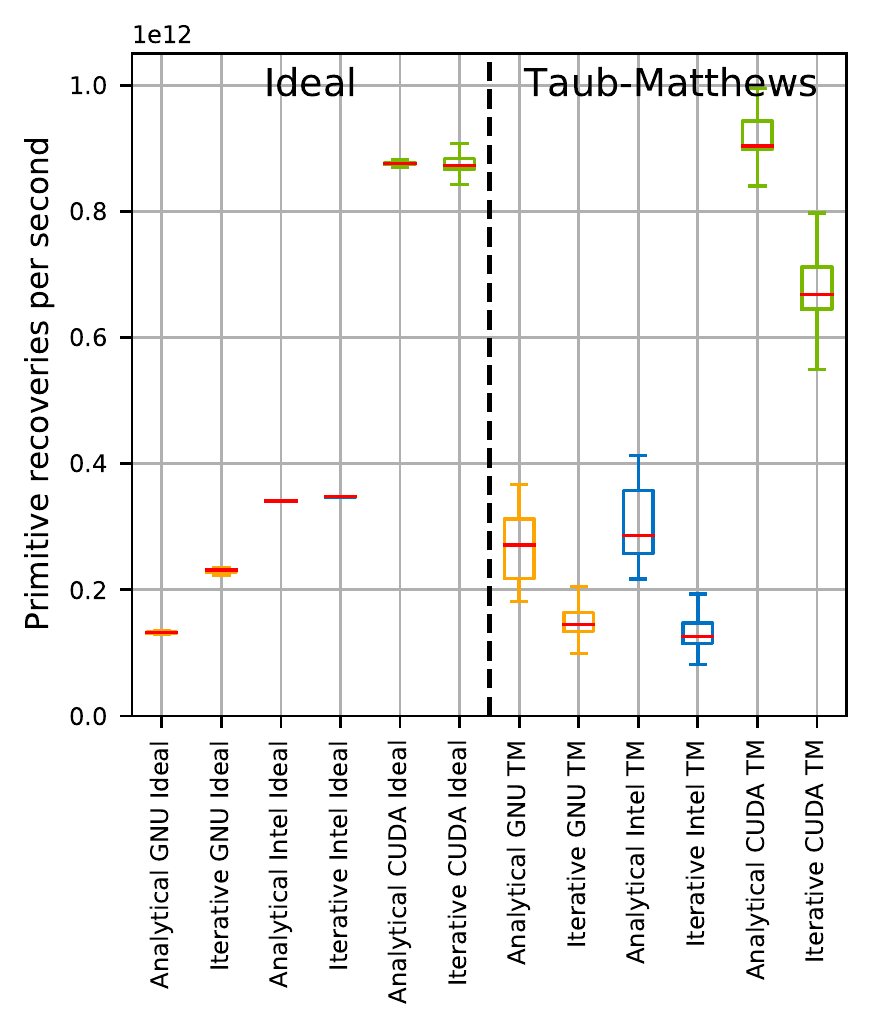}
\end{center}
  \caption{ \label{fig:conserved_to_primitive_performance} Aggregate
  performance of all methods and compilers tested shown as box and whiskers of
  the primitive recoveries per second (higher is better) across the grid of primitive states used
  in Fig.~\ref{fig:conserved_to_primitive_accuracy}. Red lines show medians,
  boxes show the interquartile range, and whiskers show the maximum and minimum
  values inside of $1.5$ times the length of the interquartile range above the
  3rd quartile and below the 1st quartile,  described by
  \cite{tukeyExploratoryDataAnalysis1977}. We exclude outlier timings from the
  figure, which range from $10^{11}$ to $1.2\times10^{12}$ primitive recoveries
  per second for all methods and compilers. We show results for GNU on CPUs in
  orange, Intel on CPUs in blue, and CUDA on GPUs in green, for the ideal gas
  on the left and the Taub-Matthews equation of state on the right. Generally,
  on CPUs using the Intel compiler allows more primitive recoveries per second
  than the GNU compiler. The performance for recovery with the Taub-Matthews gas has a
  much larger spread than recovery with the ideal equation of state. Between
  the two equations of state, the solvers achieve roughly the same number of
  recoveries per second on each architecture, indicating that equation of state
  can have a mitigated impact on the full code's performance. }
\end{figure}

In Fig.~\ref{fig:conserved_to_primitive_performance} we show performance of all
methods on all architectures and compilers tested as a box and whisker plot of
the attained primitive recoveries per second. Runs on CPUs with GNU and Intel
and the Kokkos OpenMP backend were performed on 2-socket node with Intel Xeon
Platinum 8268 CPUs on a total of 48 OpenMP threads compiled with  AVX512
vectorization.  Runs with the Kokos CUDA backend were performed on an NVidia
V100 SXM2 Tesla GPU. For the ideal gas, the analytic method is slower than the
iterative method on GNU, slightly faster on Intel, and nearly the same
performance on GPUs. For the Taub-Matthews approximation to the Taub-Matthews
equation of state, the analytical method is generally faster on all
architectures, with the performance difference being the greatest on Intel and
the smallest on GNU.  Between the two equations of states, the analytical
solver for both gases performs at about the same speed for each architecture.
This suggests that just considering conserved-to-primitive updates, using a
Taub-Matthews equation of state is about as fast as using an
ideal equation of state, although the more complex computation of wavespeeds
and enthalpies in the Taub-Matthews equation of state will lead to slowdowns
elsewhere. 

Overall, these results demonstrate that, for the ideal gas equation of state,
the iterative method to recover the primitive variables from the conserved
variables is more flexible, robust, accurate, and in some cases faster than the
analytical method. By contrast, for the Taub-Matthews equation of state, the
characteristics of the analytic and iterative solver are nearly the opposite,
with the iterative solver performing generally worse. Nevertheless, the
comparable speed and robustness of the analytical solver for the Taub-Matthews
equation of state suggest that the higher fidelity of the Taub-Matthews
equation of state comes at little cost to execution time and stability.

\section{Tests of the Relativistic Hydrodynamics Scheme}\label{sec:tests}

To verify the accuracy of the relativistic hydrodynamics scheme, we investigate
several standard test problems in 1D and 2D with and without shocks. First, in
\S\ref{sec:linearwaves}, we demonstrate convergence of a set of relativistic
linear waves in three-dimensions. We then demonstrate the accuracy of the
method for discontinuous solutions in \S\ref{sec:1d_riemann_problems} by
demonstrating convergence for five different 1D Riemann problems to high
resolution reference solutions generated from a publicly available finite
volume code \Athenapp\citep{stoneAthenaAdaptiveMesh2020}. Next, we
demonstrate the scheme's ability to handle multi-dimensional shocks through a
series of 2D Riemann problems previously established in the literature.  Then,
we measure the growth rate of the relativistic Kelvin-Helmholtz instability in
2D in \S\ref{sec:kelvin_helmholtz}, comparing to results using the finite
volume code PLUTO\citep{mignonePLUTOCODEADAPTIVE2011}. Last, in
\S\ref{sec:performance}, we show timing tests of the code evolving the
Kelvin-Helmholtz instability.


\subsection{Linear Waves} \label{sec:linearwaves}

Prior work in the literature \citep[see, e.g.][]{stoneAthenaNewCode2008} has
demonstrated that the convergence of linear waves in multi-dimensions is a
sensitive test of algorithmic fidelity. As far as we are aware, however, linear
wave convergence has not been utilized as a test of algorithms for relativistic
hydrodynamics.  Here, we elucidate how such a test can be established and
demonstrate the performance of the algorithm presented here for such a test
problem. Following the analysis of linear waves in relativistic hydrodynamics
presented in \cite{Keppens:2008}, a perturbation is made to the initial
primitive state, $\mathbf{W}_0 = [\rho_0, \mathbf{v}_0, P_0]^T$ (using rest mass density, three-velocity, and pressure), in the form
of
\begin{equation}
	\label{eq:lw_perturbation}
	\mathbf{W}[i] = \mathbf{W}_0[i] + A \mathbf{r}^j[i] \sin(k x - \omega t)
\end{equation}
where $\mathbf{W}$ is the perturbed primitive state, $A$ is the perturbation
amplitude (typically $10^{-6}-10^{-4}$), $\mathbf{r}^j[i]$ is the j$^{th}$
right eigenvector, the wavelength is equal to 1, $k=2\pi$ and $\omega = k
\lambda^j$. Here, we have defined $\lambda$ as the wavelength and $\lambda^j$
is the eigenvalue corresponding to the j$^{th}$ right eigenvector of the
Jacobian, $A(\mathbf{W})$, given in
\citet{mignonePiecewiseParabolicMethod2005}. Each eigenvalue/vector pair
corresponds to a different set of linear waves, 5 in total, which we denote with $j\in\{-,
0^{(1,2,3)}, +\}$. The initial condition in the code is specified in the conserved variables;
these are computed through symbolic computation of
\begin{equation}
	\mathbf{U}[i] (t=0) = \left.\frac{\partial \mathbf{U}}{\partial \mathbf{W}}\right\vert_{\mathbf{W}} \mathbf{W}[i] (t=0)
\end{equation}
The non-linear transformation matrix, $\partial \mathbf{U} / \partial \mathbf{W}$ must be constructed
around a state, $\mathbf{W}$ such that the solution to the non-linear
relationship $\mathbf{W}[i]](\mathbf{U}) = \mathbf{W}_0[i] + A \mathbf{r}^j[i]
\sin(k x - \omega t)$ at $t=0$. If this condition is not fulfilled, then a
\emph{different} problem is initialized and the evolution of the system will
depart from the linear dispersion relation. We have found that such a condition is fulfilled 
by constructing the non-linear transformation matrix, $\partial \mathbf{U} / \partial \mathbf{W}$ must be constructed
around a state, $\mathbf{W} = \left( \rho_0, P_0, \mathbf{v}_0, \gamma \right)$, where $\gamma = 1/\sqrt{1 - |\mathbf{v}|^2/c^2}$ includes contributions
from the perturbed velocity, $\mathbf{v} = \mathbf{v}_0[i] + A \mathbf{r}^j[i] \sin(k x - \omega t)$.

Now that the 1D perturbed states $\mathbf{U}$ and $\mathbf{W}$ have been
determined, we can rotate these for 2D and 3D non-grid-aligned cases. To do
this, we first start with a desired number of wavelengths, $N$, and find the
$n^{th}$ acceptable angle, $\theta$, by Eq.~\ref{eq:wave_angle}, where $n<N$.
The values for $N$ and $n$ for the linear waves tests are shown in
Tab.~\ref{t:N_n_values}.

\begin{equation}
	\label{eq:wave_angle}
	\theta = \tan^{-1} \left( \sqrt{\frac{N}{N-n} - 1} \right)
\end{equation}

\begin{table}[hbt!]
\caption{
	\label{t:N_n_values}
	Values of $N$ (no. of wavelengths) and $n$ ($n^{th}$ acceptable wavelength)
	for linear waves tests (see Eq.~\ref{eq:wave_angle})
}
\centering
\begin{tabular}{l|cc}
\hline
Test Type & $N$ & $n$ \\
\hline
1D & 1 & 0 \\
2D Grid-Aligned & 1 & 0 \\
2D Non-Grid-Aligned & 2 & 1 \\
3D Grid-Aligned & 1 & 0 \\
3D Non-Grid-Aligned & 3 & 2 \\
\hline
\end{tabular}
\end{table}

From here, the base equations in the 1D form of Eq.~\ref{eq:lw_perturbation}
are rotated by the angle $\theta$. Which is done either about the $y$ axis,
$a=(0,1,0)$, for 2D or about the $a=(0,-1,1)$ axis for 3D. The rotation matrix,
$\mathbf{R}$, is generated via
\begin{equation}
	\mathbf{r}_1 = \begin{bmatrix} 1 & 0 & 0 \\ 0 & 1 & 0 \\ 0 & 0 & 1 \end{bmatrix},
	\mathbf{r}_2 = \begin{bmatrix}
		a_x a_x & a_x a_y & a_x a_z \\
		a_y a_x & a_y a_y & a_y a_z \\
		a_z a_x & a_z a_y & a_z a_z \\
    \end{bmatrix},
	\mathbf{r}_3 = \begin{bmatrix}
		   0 &  -a_z &  a_y \\
		 a_z &     0 & -a_x \\
		-a_y &   a_x &    0 \\
	\end{bmatrix}
\end{equation}
\begin{equation}
	\mathbf{R} = \cos(\theta) \mathbf{r}_1 + (1-\cos(\theta)) \mathbf{r}_2 +
		\sin(\theta) \mathbf{r}_3.
\end{equation}
$\mathbf{R}$ is then used to rotate the three-velocity vector, $\mathbf{v}$, and the
momentum vector, $\mathbf{M}$, by left multiplying them by $\mathbf{R}$. Next,
the $(x,y,z)$ coordinates in each equation are substituted with rotated
coordinates $(x^{\prime}, y^{\prime}, z^{\prime})$, where
\begin{equation}
	x^{\prime} = \mathbf{R} \begin{bmatrix} 1 \\ 0 \\ 0 \end{bmatrix}, \quad
	y^{\prime} = \mathbf{R} \begin{bmatrix} 0 \\ 1 \\ 0 \end{bmatrix}, \quad
	z^{\prime} = \mathbf{R} \begin{bmatrix} 0 \\ 0 \\ 1 \end{bmatrix}.
\end{equation}
Once these values have been substituted, the final, non-grid-aligned equations
for $\mathbf{U}$ and $\mathbf{W}$ have been obtained.

For all eigenvalue/eigenvector cases, $j=\{-, 0^{(1,2,3)}, +\}$, tests are run for
the rotation configurations in Table \ref{t:N_n_values} with basis order and
time integrator combinations of (0, RK1), (1, SSPRK2), and (2, SSPRK3). The
domain, $\mathbf{L}$, and number of elements in each direction, $\mathbf{N}$,
is calculated based on the rotation matrix,
$\mathbf{R}$:
\begin{equation}
	\mathbf{L} = N \mathbf{R} \left( \frac{\mathbf{e}}{|\mathbf{e}|} \right)
\end{equation}
\begin{equation}
  \mathbf{N} = N n_{\text{elem}} x_{\sigma}^r \mathbf{R} \left( \frac{\mathbf{e}}{|\mathbf{e}|}
	\right)
\end{equation}
where $N$ is the number of wavelengths, $\mathbf{e}$ is the direction vector
for the default orientation of the wave $\left( \begin{bmatrix} 1 & 0 & 0
\end{bmatrix}^T \right)$, $x_{\sigma}$ is the refinement multiplier per refinement
increment (default $x_{\sigma}=2$), $r$ is the refinement level, and $n_{\text{elem}}$
is the base number of elements, which varies for 1D, 2D, and 3D.

\begin{figure}[hbt!]
\centering
\subcaptionbox{Case: -}        [.32\columnwidth]{\includegraphics[width=0.32\columnwidth]{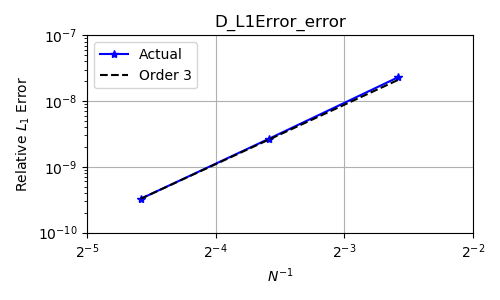}}
\subcaptionbox{Case: $0^{(1)}$}[.32\columnwidth]{\includegraphics[width=0.32\columnwidth]{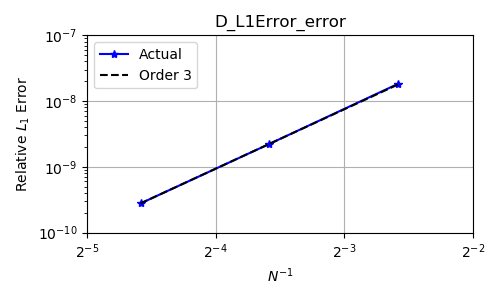}}
\subcaptionbox{Case: $0^{(2)}$}[.32\columnwidth]{\includegraphics[width=0.32\columnwidth]{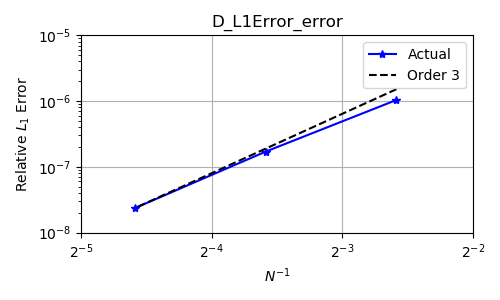}}
\subcaptionbox{Case: $0^{(3)}$}[.32\columnwidth]{\includegraphics[width=0.32\columnwidth]{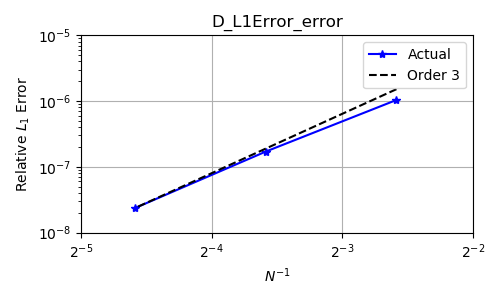}}
\subcaptionbox{Case: +}        [.32\columnwidth]{\includegraphics[width=0.32\columnwidth]{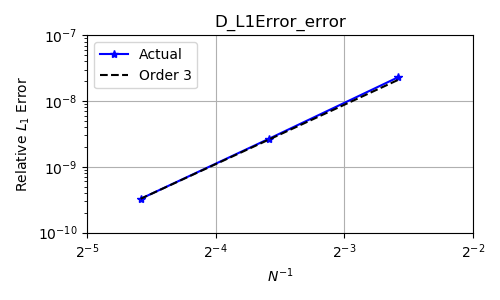}}
  \caption{Order of convergence for the relativistic mass density (in solid
  blue) for three resolutions along the $x$-axis the 5 eigenvalue/eigenvector
  pairs $j\in\{-,0^{(1,2,3)}, +\}$ in different panel. For all tests here we
  test in 3D with non-grid-aligned waves, using a $2^{\text{nd}}$ order basis
  with the SSPRK3 integrator. For all cases we expect a $3.0$ rate of
  convergence, which we denote with a dashed black line.}
\label{fig:linear_waves_errors}
\end{figure}

For these tests, the velocity was either set to $\mathbf{v} =
\mathbf{0}$ or $\mathbf{v} = \begin{bmatrix} 0.5v_{\text{max}} &
-0.3v_{\text{max}} & 0.4v_{\text{max}} \end{bmatrix}^T$, where $v_{\text{max}}
= 0.05 c_s$.  The base time step is determined by running the test with
adaptive time stepping, which adjusts the time step to maintain a certain CFL
during the test (0.2 in this case).  The test is then run again 3 times, each
time increasing the refinement in both space and time by a factor of 2 to
maintain a constant CFL. The L1Error and L2Error are gathered for each test and
are fitted against the results using the following equation:
\begin{equation}
  \text{L1Error}(dx) = p_0 + p_1(dx)^{p_2}
\end{equation}
where $p_0$, $p_1$, and $p_2$ are fitting constants.  The exponent $p_2$ is the
convergence order, which is expected to be 1, 2, and 3 for the time integrators
RK1, SSPRK2, and SSPRK3 respectively. Results for the 3D,
non-grid-aligned, zero velocity, basis order 2, SSPRK3, test case are shown in
Tab.~\ref{t:linear_waves_convergence_orders}, while the L1Error is plotted
against the expected values for the conserved quantity $D$ in
Fig.~\ref{fig:linear_waves_errors}.

\begin{table}[hbt!]
\caption{
	\label{t:linear_waves_convergence_orders}
  Order of convergence for both primitive and conserved variables along the rows for each of
  the 5 eigenvalue/eigenvector pairs $j\in\{-,0^{(1,2,3)}, +\}$ along the columns, all tested in 3D
  with non-grid-aligned waves, using a $2^{\text{nd}}$ order basis with the
  SSPRK3 integrator. For all cases we expect a $3.0$ rate of convergence.
  Entries with '-' denote variables where the eignvector used for that test
  does not affect that variable.
}
\centering
\begin{tabular}{c|ccccc}
\hline
\multirow{2}{*}{Quantity} & \multicolumn{5}{c}{Eigenvalue/eigenvector Test Case} \\
         &        - & $0^{(1)}$ & $0^{(2)}$ & $0^{(3)}$ &        + \\
\hline
     $D$ & 3.099989 &  3.036570 &  2.561624 &  2.561624 & 3.099989 \\
   $M_x$ & 3.079648 &         - &  2.838988 &  2.838988 & 3.079648 \\
   $M_y$ & 3.079648 &         - &  2.879077 &  2.824568 & 3.079648 \\
   $M_z$ & 3.079648 &         - &  2.824568 &  2.879077 & 3.079648 \\
     $E$ & 3.099989 &  3.036570 &  2.561652 &  2.561652 & 3.099989 \\
  $\rho$ & 3.099989 &  3.036570 &         - &         - & 3.099989 \\
   $u_x$ & 3.079655 &         - &  2.838988 &  2.838988 & 3.079655 \\
   $u_y$ & 3.079655 &         - &  2.879077 &  2.824568 & 3.079655 \\
   $u_z$ & 3.079655 &         - &  2.824568 &  2.879077 & 3.079655 \\
     $P$ & 3.099989 &         - &         - &         - & 3.099989 \\
\hline
\end{tabular}
\end{table}

\subsection{1D Riemann Problems}\label{sec:1d_riemann_problems}

We now investigate the accuracy of the relativistic hydrodynamics method
through considering the evolution of a set of standard 1D Riemann problems in
order to characterize how well the code handles shocks. For initial conditions,
we use three standard blast waves and a reflecting wall test from
\citet{martiNumericalHydrodynamicsSpecial2003,martiGridbasedMethodsRelativistic2015}
and one Sod shock tube, and a reflecting wall test for a total of five
different 1D Riemann problems. 

For the first four 1D Riemann problems, we use a $[0,1]$ grid with Dirichlet
boundary conditions. These four tests begin divided into a primitive state on
the left $\mathbf{W}_L = (\rho,v_x,v_y,p)_L$ for $x\in[0,0.5)$ and right
$\mathbf{W}_R = (\rho,v_x,v_y,p)_R$ for $x\in[0.5,1]$. In the fifth
test, we replace the boundary condition at $x=1$ with a reflecting boundary and
use a uniform initial primitive state through the domain. In all cases, we set
$v_z=0$ and use the ideal equation of state with $\gamma=5/3$ for the first four
tests and $\gamma=4/3$ for the fifth test.

For each of the five 1D Riemann problems, we use a $[0,1]$ grid with Dirichlet
boundary conditions except for test 5, which uses a reflecting boundary
condition on the right wall. The tests begin divided into a primitive state on
the left $\mathbf{W}_L = (\rho,v_x,v_y,p)_L$ for $x\in[0,0.5)$ and right
$\mathbf{W}_R = (\rho,v_x,v_y,p)_R$ for $x\in[0.5,1]$ except for test 5, which
begins with a constant primitive state throughout the volume. In all cases, $v_z=0$.

\begin{figure}[hp!]
  \includegraphics[width=\linewidth]{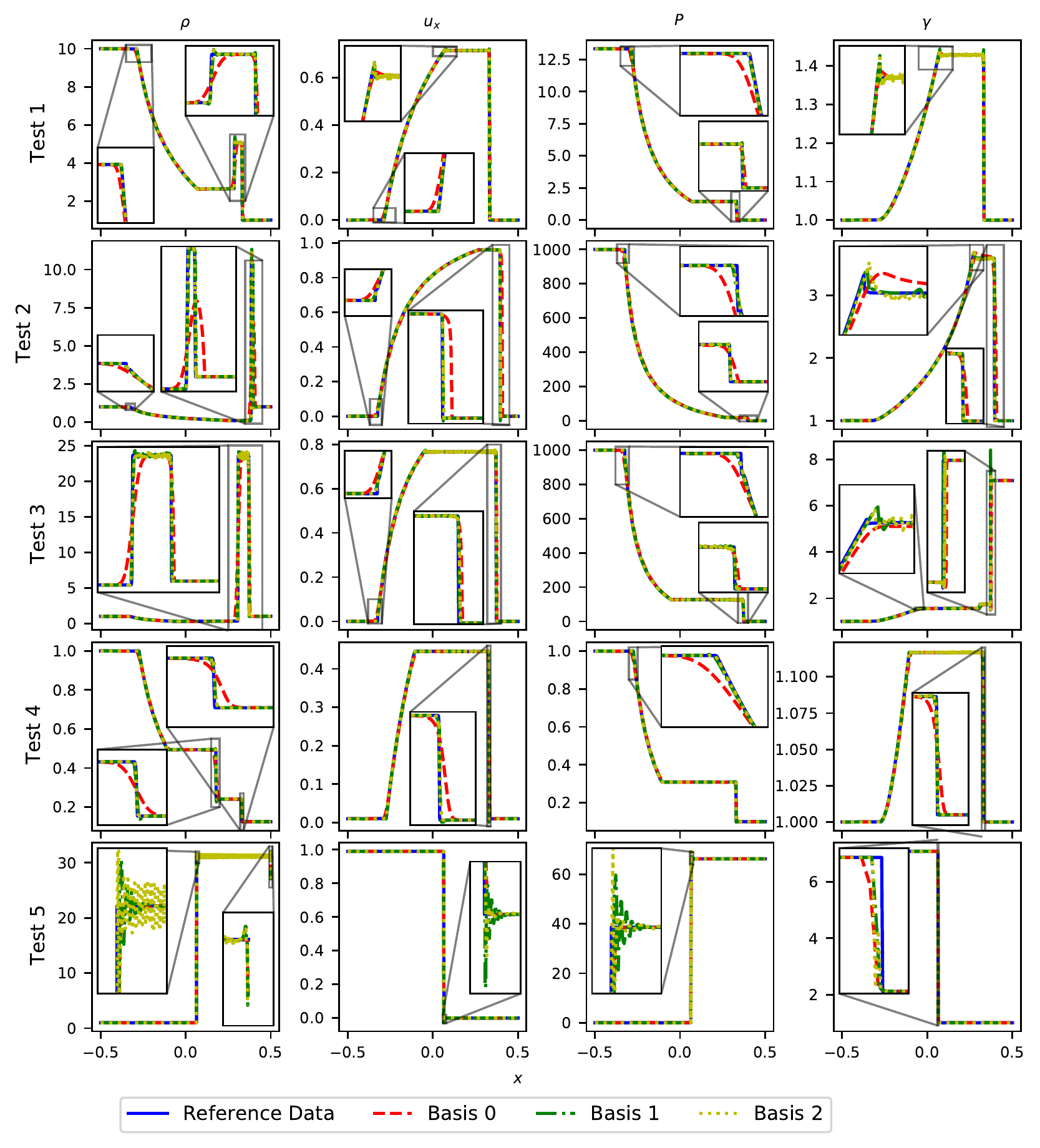}
  \caption{
    \label{fig:1d_shock}
    Plots of the five 1D Riemann problems tested using the ideal equation of
    state. Each row shows end state of a different Riemann problem. From top to
    bottom, the first row shows a mildly relativistic blast wave, the second  a
    highly relativistic blast wave, the third a blast wave with transverse
    velocity, the fourth a Sod shock tube, and the fifth a planar shock
    reflection. The columns show from left to right the rest-mass density, the
    pressure, the velocity, and the Lorentz factor.  In each panel we show the
    reference solution computed with a finite volume scheme
    \citep{stoneAthenaAdaptiveMesh2020} with a solid line and the basis 0,
    1, and 2 solutions with our method with a red dashed, green dot-dashed, and
    yellow finely dash line respectively. Although the method can evolve these
    shocks with the help of the physicality-enforcing operator, small
    oscillations appear around shocks for higher order bases. These oscillations can
    be damped out by widening the limiting thresholds for the Moe limiter or by
    changing the minmod limiter but this results in more diffusion and lower
    order convergence for basis order 2.
  }
\end{figure}

\begin{figure}[hbt!]
  \includegraphics[width=\linewidth]{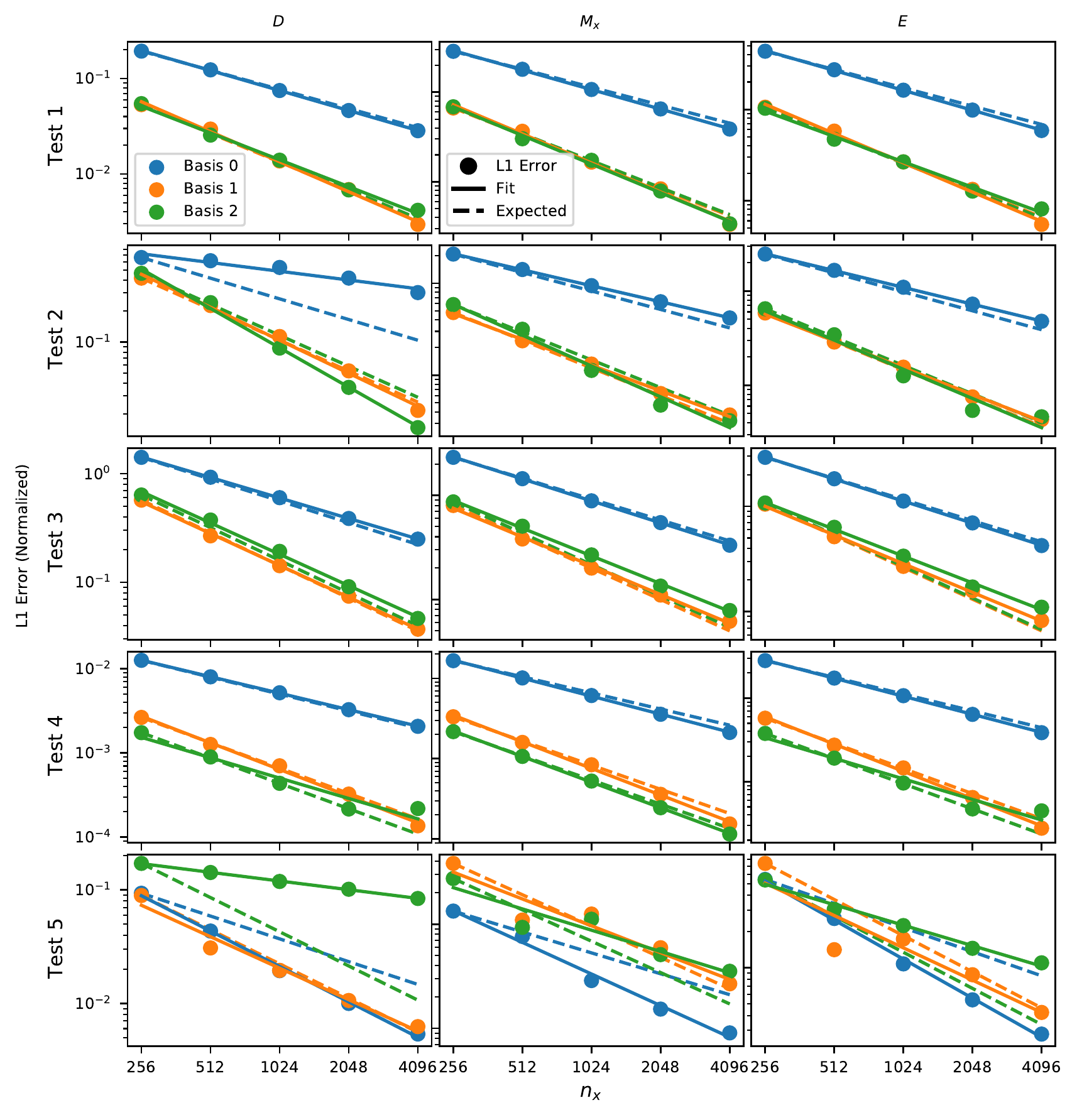}
  \caption{
    \label{fig:1d_shock_convergence}
    Convergence of the L1 error of the method presented here to a high resolution reference solution
    of the same Riemann problems from Fig.~\ref{fig:1d_shock} computed with a finite volume scheme
    \citep{stoneAthenaAdaptiveMesh2020}.
    From top to bottom, the first row shows a mildly relativistic blast wave,
    the second  a highly relativistic blast wave, the third a blast wave with
    transverse velocity, the fourth a planar shock reflection, and the fifth a
    Sod shock tube. The columns show from left to right the rest-mass density,
    the pressure, the velocity, and the Lorentz factor. In each panel we show
    the L1 error of our method with dots, a fitted convergence rate using
    logarithmically weighted least squares with a solid line, and a $2/3$
    convergence rate for basis order 0 and a first order convergence rate for
    bases 1 and 2 with dashed lines. We use different colors to denote
    different basis orders, using blue for basis order 0, orange for basis
    order 1, and green for basis order 2. Due to the presence of shocks, we
    expect the L1 error of higher order bases to converge to first order at
    best, although sharp blasts prove difficult for convergence.
  }
\end{figure}

For reference data, we compute a $n_x=2^{14}$ cell solution using a HLLC
Riemann solver, a second order Van-Leer integrator due to
\citet{stoneAthenaAdaptiveMesh2020} for each of the tested Riemann
problems.  We run the each 1D Riemann problem with five resolutions in powers
of two from $n_x=256$ to $n_x=4096$ cells with polynomial basis orders $0$,
$1$, and $2$ using the HLLC Riemann solver and the iterative primitives
recovery method for the ideal gas. For basis orders $1$ and $2$, we use the
limiter from \citet{moeSimpleEffectiveHighOrder2015} in addition to the
physicality-enforcing operator from
\S~\ref{sec:physicality_enforcing_operator}. The physicality-enforcing operator
was necessary for all tests with basis orders over $0$.
Fig.~\ref{fig:1d_shock} shows the density, longitudinal velocity, pressure, and
Lorentz factor from the five 1D Riemann problems using $n_x=128$ with the three
polynomial basis orders and the reference solution.
Fig.~\ref{fig:1d_shock_convergence} shows a log-log plot of the L1 error of the
relativistic density, longitudinal relativistic momentum density, and total
energy density compared to the reference solution along with power fits to the
convergence rate and the expected rate of convergence.

1D Riemann problem 1 is a mildy relativistic blast wave with  initial
conditions \begin{equation} \mathbf{W}_L = \left ( 10, 0, 0, (40/3)c^2
\right)_L \qquad \mathbf{W}_R = \left ( 1, 0, 0, (2/3 \times 10^{-6})c^2 \right
)_R \end{equation} where we have followed \citet{nunez-delarosaHybridDGFV2018}
and used a pressure close to zero for the right side primitive state for
numerical reasons. For this test, we use an adiabatic index $\Gamma=5/3$. We
evolve the shock until $t=0.4/c$. For this first test we achieved the expected
convergence rate in all variables except for the density for basis order 0,
which suffers from slow converging dissipation around the blast wave. We also
see a small cusp in velocity and oscillations in basis order 2 at the trailing
edge of the blast wave which are more apparent in the Lorentz factor. L1 error
of basis orders 1 and 2 are comparable, highlighting the difficultly in
achieving high-order convergence with higher order methods when the problem
contains shocks. However, since the basis order 2 test has more degrees of
freedom than the basis order 1 test, the L1 error per degree of freedom is
still lower for basis order 2, indicating that higher order bases can still be
more efficient.

1D Riemann problem 2 is a highly relativistic blast wave with initial conditions
\begin{equation}
  \mathbf{W}_L = \left ( 1, 0, 0, (10^3)c^2 \right )_L \qquad
  \mathbf{W}_R = \left ( 1, 0, 0, (10^{-2})c^2 \right )_R,
\end{equation}
using an adiabatic index $\Gamma=5/3$ and evolved until $t=0.4/c$. In this
test, we see that the sharpness of the resolved density of the blast wave
changes with resolution. We see it the sharpest with basis order 1, second with
basis order 0, and most diffuse with basis order 2, although for each basis the
sharpness improves with resolution. We see a slight cusp in the Lorentz factor
for all basis orders just behind the blastwave where the velocity approaches
$c$ but in the high resolution finite volume method the region has a flat
Lorentz factor. The sharp blast wave in density causes problems for convergence
at basis order 0 while higher order bases achieve the expected convergence.

1D Riemann problem 3 is also a highly relativistic blast wave but with a
transverse velocity with initial conditions
\begin{equation}
  \mathbf{W}_L = \left ( 1, 0, 0, (10^3)c^2 \right )_L \qquad
  \mathbf{W}_R = \left ( 1, 0, 0.99, (10^{-2})c^2 \right )_R,
\end{equation}
with an adiabatic index $\Gamma=5/3$ and evolved until $t=0.4/c$. With the
addition of a relativistic transverse velocity, the blast wave widens into a
square plateau in density, somewhat similar to problem 1. Like in problem 2, we
find that basis order 1 best captures the blast wave, although resolution
improves accuracy for all basis orders. In the Lorentz factor we see a small
cusp at the rightmost edge of the rarefaction and some smearing across the
blastwave. The wider blast wave allows basis order 0 to achieve the expected
convergence rate. L1 error for basis order 2 is greater than the L1 error for
basis order 1, although this is mostly due to more degrees of freedom in the
summation of the L1 error for basis order 1.

1D Riemann problem 4 is a Sod shock with initial conditions
\begin{equation}
  \mathbf{W}_L = \left ( 1, 0.01 c, 0, 1.0 c^2 \right )_L \qquad
  \mathbf{W}_R = \left ( 0.125, 0.01 c, 0, 0.1 c^2 \right )_R,
\end{equation}
using an adiabatic index $\Gamma=4/3$ and evolving until $t=0.4/c$. We see some
diffusivity across the contact discontinuity and at the leftmost edge of the
rarefaction.

For the fifth 1D Riemann problem we study a highly relativistic flow moving to the right and reflecting
against the right wall.  We use the initial conditions
\begin{equation}
  \mathbf{W} = \left ( 1, 0.99999 c, 0, 0.01 c^2 \right ),
\end{equation}
with an adiabatic index $\Gamma=4/3$ and evolved until $t=1.5/c$. We see a small
cusp in the Lorentz factor at the left edge of the piled up stationary mass.
For higher order bases, we see wall heating causing spurious oscillations in
the reflected fluid. These leads to slow rates of convergence for basis order
2.


\subsection{1D Taub-Matthews Equation of State Test}\label{sec:1d_synge_gas_test}

We test the Taub-Matthews approximation to the Synge equation of state against
the ideal equation of state using the fifth blast wave problem from
\citet{ryuEquationStateNumerical2006}, which highlights the differences between
the Synge gas and ideal gas.  The initial conditions for the test, using the
same notation and domain as
\S\ref{sec:1d_riemann_problems}, are
\begin{equation}
  \mathbf{W}_L = \left ( 1, 0, 0.9  c, (10^3)c^2 \right )_L \qquad
  \mathbf{W}_R = \left ( 1, 0, 0.99 c, (10^{-2})c^2 \right )_R,
\end{equation}
which evolves into a blast wave.  In the initial state, the temperature
stand-in $\Theta=P/\rho$ on the left-hand side is relativistic while $\Theta$
on the right-hand side is non-relativistic. As such, for an ideal equation of
state, an adiabatic index of $\Gamma=4/3$ is appropriate for the left-hand side
while $\Gamma=5/3$ is appropriate for the right-hand side. The Taub-Matthew
equation of state approximation allows accurate modeling of both sides with a
single equation of state.

\begin{figure}[hbt!]
  \centering
  \includegraphics[]{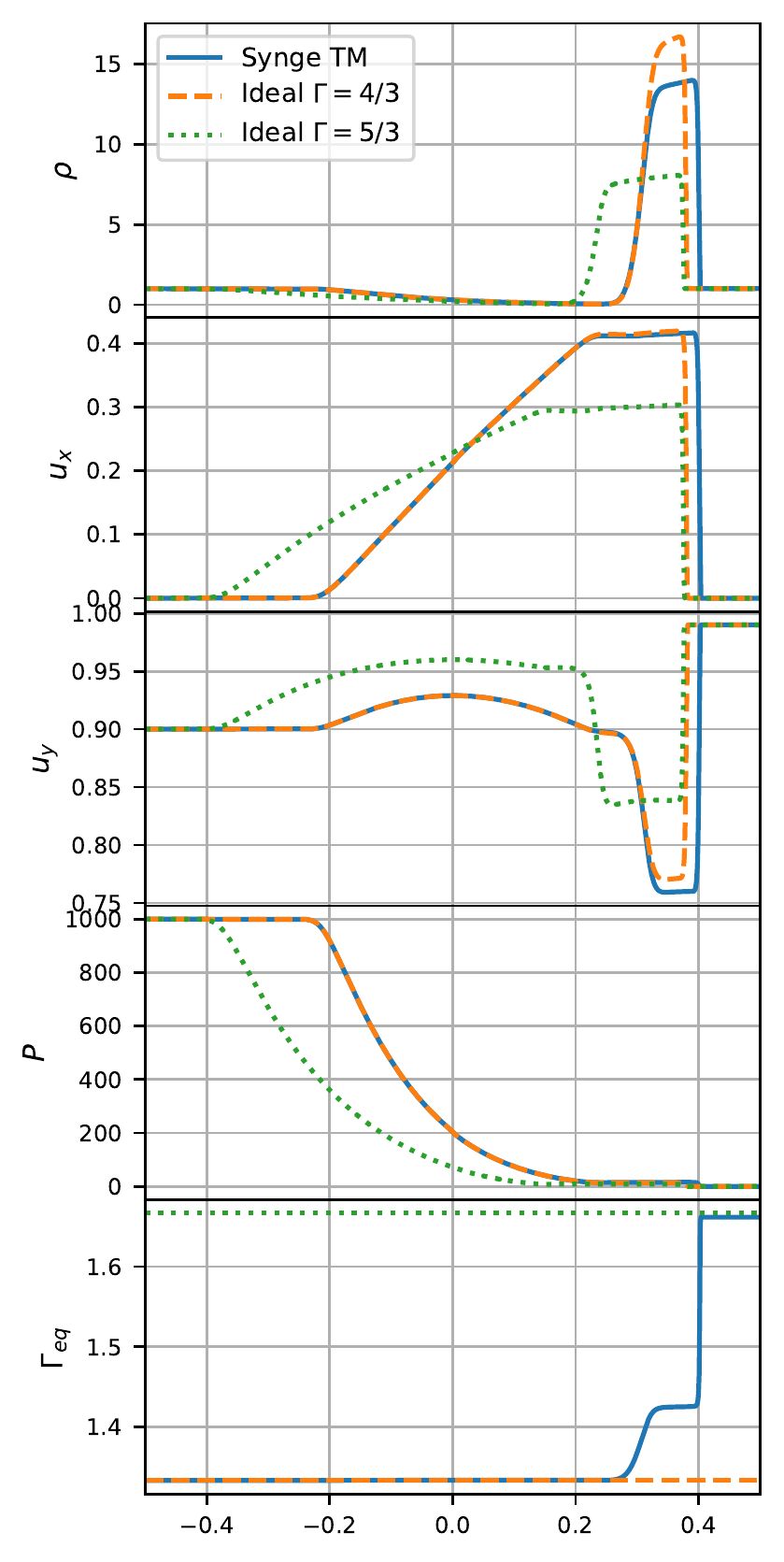}
  \caption{
    \label{fig:synge_blast_wave}
    Blast wave with relativistic temperatures on the left and non-relativistic
    temperature on the right, evolved to $t=0.7$ using the Taub-Matthews equation of
    state (solid blue), ideal equation of state with adiabatic index
    $\Gamma=4/3$ (dashed orange), and ideal equation of state with $\Gamma=5/3$
    (finely dashed green). In order of rows, we show the density $\rho$,
    longitudinal velocity $u_x$, transverse velocity $u_y$, pressure $P$, and
    equivalent adiabatic index $\Gamma_{\text{eq}} =\left (  h - c^2\right) /\left(h - c^4 -
    P/\rho \right)$. The Taub-Matthews equation of state, as an approximation
    to the Synge gas, behaves apart from both the $\Gamma=5/3$ and $\Gamma=4/3$
    ideal gases depending on the effective adiabatic index.
  }
\end{figure}

We show results for the blast wave with the three different equation of state
in Fig.~\ref{fig:synge_blast_wave}. The Synge gas as approximated by the
Taub-Matthews equation of state behaves like the relativistic $\Gamma=4/3$
ideal gas on the left side of the blast wave (which is contained within
$[0.3,0.4]$ at $t=0.7$ as shown) and like the non-relativistic $\Gamma=5/3$
ideal gas on the right side.  This is most evident in the velocity profiles and
pressure profiles in the relativistic region that occupies most of the domain at
this time. The equivalent adiabatic index $\Gamma_{\text{eq}}$ of the
Taub-Matthews equation of state is expectedly $4/3$ in the relativistic region
and $5/3$ in the non-relativistic region, and varies between these values
across the blast wave.  In this region within the blast wave, the peak density
with the Taub-Matthews equation of state falls between the extremes of the two
ideal gases. Notably, the blast wave with the Taub-Matthew equation of state
travels slightly faster than either ideal gases, and the minimum transverse
velocity is also lower. These results are consistent with the blast waves
evolved with the Taub-Matthews equation of state in
\citet{ryuEquationStateNumerical2006}.

\subsection{2D Riemann Problems} \label{sec:2d_riemann_problems}

Next, we test the robustness of the method evolving intersecting shocks in 2D
using the three 2D Riemann problems used in
\citet{zannaEfficientShockcapturingCentraltype2002,nunez-delarosaHybridDGFV2018}.
In each of the three problems, the problem is defined with a
$[-1,1]\times[-1,1]$ domain divided into four quadrants with different initial
states. Following \citet{nunez-delarosaHybridDGFV2018}, we denote these states
using
\begin{align}
  \mathcal{Q}_1 & := [0,1] \times [0,1]\\
  \mathcal{Q}_2 & := [-1,0] \times [0,1]\\
  \mathcal{Q}_3 & := [-1,0] \times [-1,0]\\
  \mathcal{Q}_4 & := [0,1] \times [-1,0]
\end{align}
and denote the initial primitive states in each of these quadrants by
$\mathbf{W}_1$, $\mathbf{W}_2$, $\mathbf{W}_3$, and $\mathbf{W}_4$
respectively. For all of these Riemann problems, we use an adiabatic index of
$\Gamma=5/3$, use $v_z=0$ everywhere, and  use transmissive boundary conditions
on all sides. We evolve each Riemann problem to $t=0.8/c$. For all 2D shock tests we use
the Moe limiter \citep{moeSimpleEffectiveHighOrder2015} and HLLC Riemann solver.

\subsubsection{ 2D Riemann Problems: Test 1} \label{sec:2d_riemann_problems_test_5}

In this test, the domain begins with a low density and pressure region in the
upper right, a high density and pressure region in the lower left, and
intermediate density and high pressure regions in the upper left and lower
right with initial velocities moving into the lower density region with
$\beta=0.7$.
\begin{align}
  \label{eq:2d_riemann_test5}
  \mathbf{W}_1 & :=  ( 0.035145216124503, 0.0, 0.0, 0.162931056509027 c^2) \\
  \mathbf{W}_2 & :=  (0.1, 0.7 c, 0.0, 1.0 c^2) \\
  \mathbf{W}_3 & :=  (0.5, 0.0, 0.0, 1.0 c^2) \\
  \mathbf{W}_4 & :=  (0.1, 0.0, 0.7 c, 1.0 c^2)
\end{align}
\begin{figure}[hbt!]
  \includegraphics[width=\linewidth]{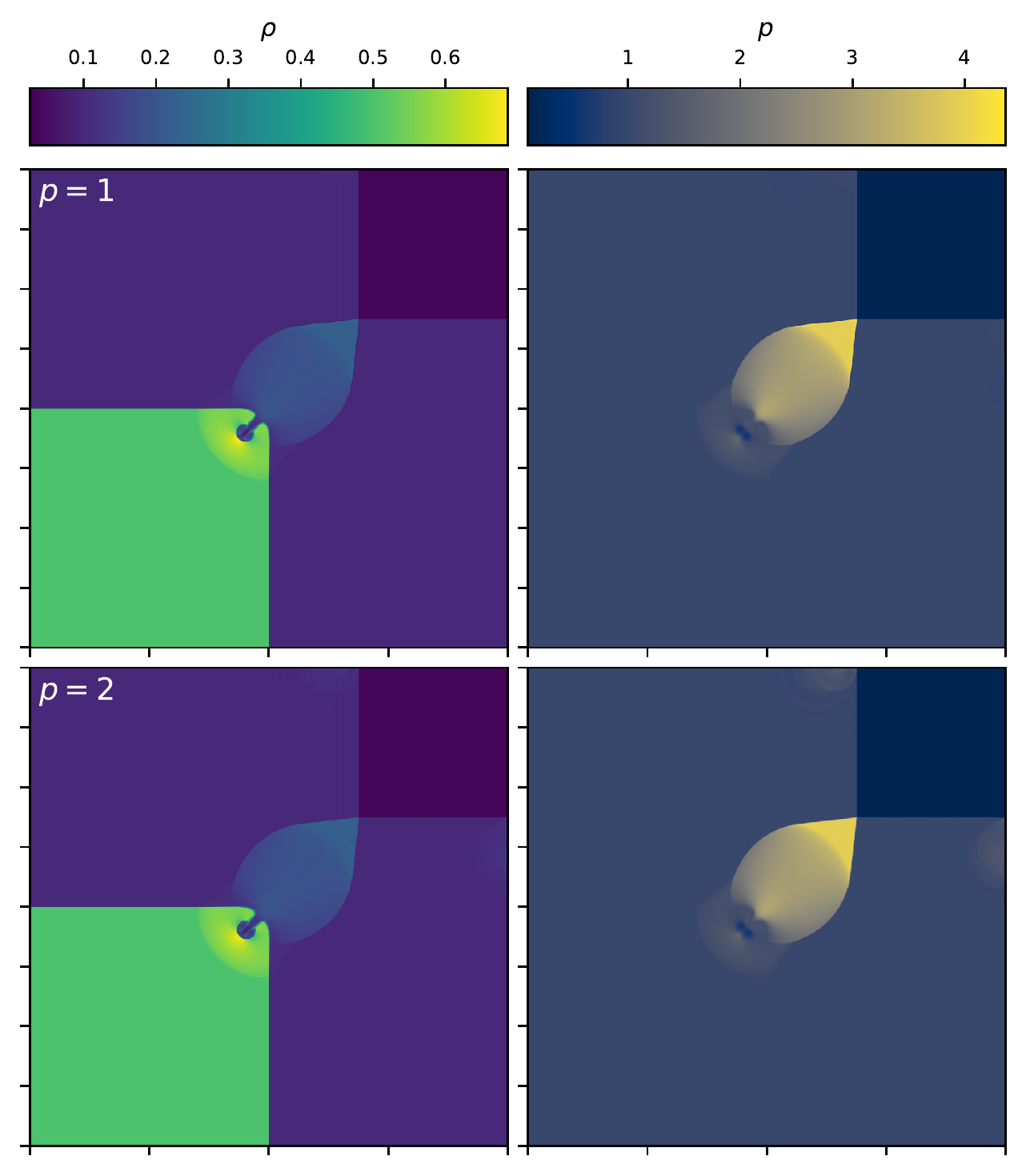}
  \caption{
    \label{fig:2d_riemann_plot_5}
    Plots of the 2D Riemann problem test 1 with two colliding shocks using the
    initial conditions in eq. \ref{eq:2d_riemann_test5}, using a
    $1^{\text{st}}$ order basis in the top row and a $2^{\text{nd}}$ order
    basis in the bottom row. We show the rest-mass density in the left column
    and the pressure in the right column at $t=0.8/c$ on a grid with
    $1024\time1024$ elements.  Note the boundary effects where shocks traveling
    into the first quadrant intersect with the outflow boundaries when  using
    the $2^{\text{nd}}$  order basis.
  }
\end{figure}

Results from the first 2D Riemannn problem is shown in
Fig.~\ref{fig:2d_riemann_plot_5} with the $1^{\text{st}}$ and $2^{\text{nd}}$
order bases, the system evolves with stationary contact discontinuities between
the high density and moving intermediate density regions, planar shocks moving
from the intermediate density regions into the low density regions, and curved
shocks bowing into the intermediate density regions from the diagonal. A
jet-like, low density structure forms into the high density region with gentle
density and pressure gradients forming ahead and behind it. Our method evolves
the curved shocks with symmetric shock fronts using both low order and
high-order bases. When using bases over $0^{\text{th}}$ order, the
physicality-enforcing operator described in
\S\ref{sec:physicality_enforcing_operator} is necessary to avoid negative
densities, pressures, and otherwise unphysical states. With the $2^{\text{nd}}$
order basis, we see subtle boundary effects where the shocks traveling
transverse to the boundary into the first quadrant intersect with  the outflow
boundary conditions. Boundary effects with the $2^{\text{nd}}$ order basis are
seen again in \S~\ref{sec:2d_riemann_problems_test_6} and
\S~\ref{sec:kh_non_linear}.

\subsubsection{ 2D Riemann Problems: Test 2} \label{sec:2d_riemann_problems_test_6}

In this test, all four quadrants begin with different densities, equal
pressures, and each move diagonally clockwise around the origin.

\begin{align}
  \label{eq:2d_riemann_test6}
  \mathbf{W}_1 & :=  (0.5, 0.5 c, -0.5 c, 5.0 c^2) \\
  \mathbf{W}_2 & :=  (1.0, 0.5 c, 0.5 c, 5.0 c^2) \\
  \mathbf{W}_3 & :=  (3.0, -0.5 c, 0.5 c, 5.0 c^2) \\
  \mathbf{W}_4 & :=  (1.5, -0.5 c, -0.5 c, 5.0 c^2)
\end{align}

\begin{figure}[H]
  \includegraphics[width=\linewidth]{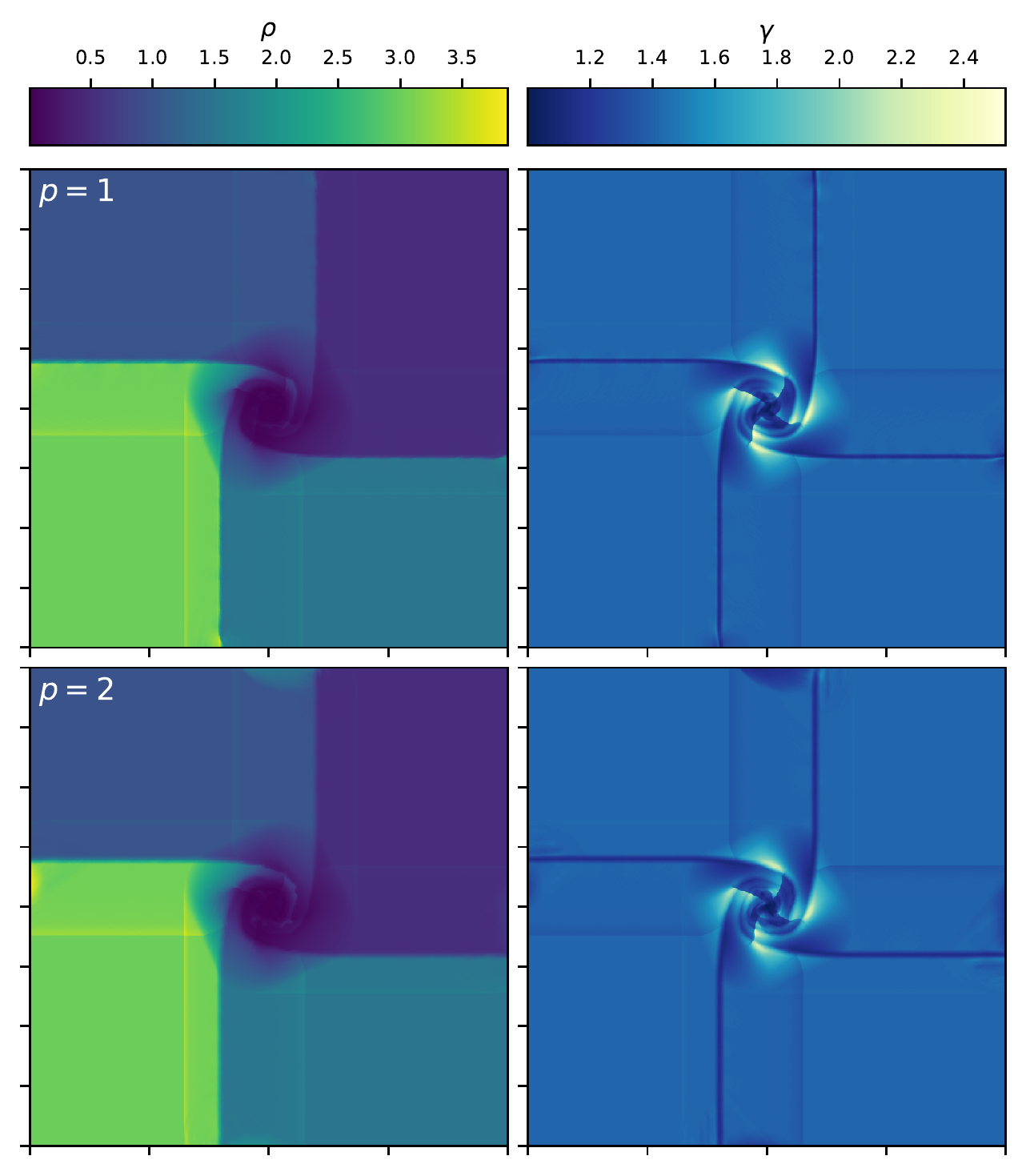}
  \caption{
    \label{fig:2d_riemann_plot_6}
    Plots of the 2D Riemann problem test 2 with four vortex sheets using the
    initial conditions in eq. \ref{eq:2d_riemann_test6}, using $1^{\text{st}}$
    order basis in the top row and a $2^{\text{nd}}$ order basis in the bottom
    row. We show the rest-mass density in the left column and the pressure in
    the right column at $t=0.8/c$ using a grid with $1024\time1024$ elements.
    Note the boundary effects where the vortex sheets intersect with the
    outflow boundaries which are subtle using the $1^{\text{st}}$ order basis
    and more apparent when using the $2^{\text{nd}}$ order basis, especially
    along the top boundary. Like the 1D test of a shock reflecting against a
    wall, this test highlights unresolved difficulties of higher order bases
    leading to boundary effects.
  }
\end{figure}

Results from the second 2D Riemannn problem are shown in
Fig.~\ref{fig:2d_riemann_plot_6} with the $1^{\text{st}}$ and $2^{\text{nd}}$
order bases, the system develops into four vortex sheets that expand from the
origin. A low rest mass region forms at the center of the vortex sheets at the
origin. The physicality-enforcing operator ensures positive densities and
pressures in this region.  With the $2^{\text{nd}}$ order basis, we see subtle
boundary effects where the shocks traveling transverse to the boundary into the
first quadrant intersect with  the outflow boundary conditions. These boundary
effects are not apparent with the $1^{\text{st}}$ order basis.

\subsubsection{ 2D Riemann Problems: Test 3} \label{sec:2d_riemann_problems_test_7}

This tests begins with overdense first and third quadrants following
\begin{align}
  \label{eq:2d_riemann_test7}
  \mathbf{W}_1 & :=  (1.0, 0.0, 0.0, 1.0 c^2) \\
  \mathbf{W}_2 & :=  (0.5771, -0.3529 c, 0.0, 0.4 c^2) \\
  \mathbf{W}_3 & :=  (1.0, -0.3529 c, -0.3529 c, 1.0 c^2) \\
  \mathbf{W}_4 & :=  (0.5771, 0.0, -0.3529 c, 0.4 c^2).
\end{align}
Rarefactions move from the second and fourth quadrants into the first and third
quadrants, producing curved shocks where the rarefactions intersect.

\begin{figure}[hbt!]
  \includegraphics[width=\linewidth]{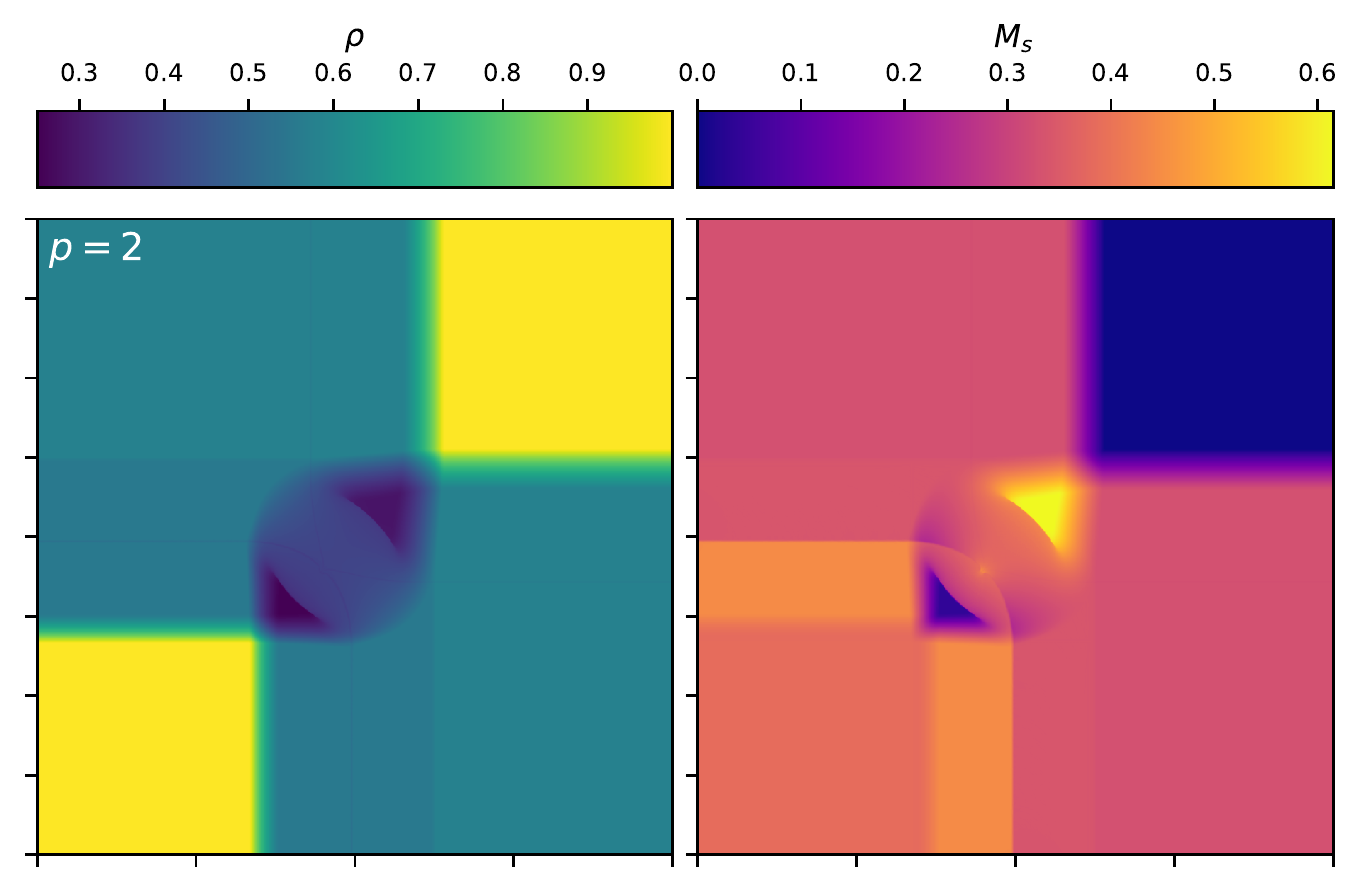}
  \caption{
    \label{fig:2d_riemann_plot_7}
    Plots of the 2D Riemann problem test 3 with intersecting rarefactions  using the
    initial conditions in eq. \ref{eq:2d_riemann_test7}. We show the rest-mass
    density left column and the pressure right column at $t=0.8/c$
    using a $2^{\text{nd}}$ order basis on a grid with $1024\time1024$ elements.
  }
\end{figure}

Results from the third 2D Riemann problem are shown in
Fig.~\ref{fig:2d_riemann_plot_7} with the $2^{\text{nd}}$ order basis. The
method evolves the curved shocks and rarefactions without issue. No boundary
effects are apparent in this test.


\subsection{Kelvin-Helmholtz Instability} \label{sec:kelvin_helmholtz}

The relativistic Kelvin-Helmholtz instability provides a useful benchmark with which
to explore the performance of the scheme presented here for shear-flow type problems.
Previous work, e.g. \cite{Mignone:2009,Beckwith:2011} has revealed signficant differences
in the performance of different numerical schemes for this classic fluid flow problem and
subsequent work \cite{Lecoanet:2016} has further elucidated the issues raised in prior
works through the comparison of finite volume and spectral methods. Here, we compare
the discontinuous-Galerkin scheme presented here with a finite volume method previously
presented in the literature \cite{mignonePLUTOCODEADAPTIVE2011}, explore both the linear and non-linear
regime of the instability and examine performance metrics for the scheme.

We simulate the Kelvin-Helmholtz instability on a $[-0.5,0.5]\times[-1.0,1.0]$
domain with a single interface along $y=0$, specified with a smoothly varying profile
using a mesh of square cells with twice as many cells in $y$ than $x$, testing mesh sizes in powers of $2$ from
$256\times512$ to $4096\times8192$ for a total of $6$ different mesh sizes.  We
tested using basis orders $0$, $1$, and $2$, however due to memory constraints
and increasing execution time, we forgo the highest resolution mesh using basis
order $1$ and the two highest resolutions using basis order $2$.  We conduct
separate tests using the HLLC and HLL Riemann solvers and using a shear
velocity $v_{x,0}=0.25c$.  We run a total of $60$ simulations
exploring growth rates of the Kelvin-Helmholtz instability. In all these calculations,
we use an ideal equation of state with adiabatic index $\gamma=4/3$ using the
iterative conserved-to-primitive solver, an initial density $\rho_0=1$, an
initial pressure $P_0=c^2$, a perturbation amplitude $A=0.05$, and a shearing
layer thickness $a=0.01$. We use $k=2\pi$ so that the wavelength of the
perturbations in $x$ is $1$ and for each test run until $t=5$ to verify from the
growth rate that the transverse velocity perturbations have saturated past the
linear growth phase.

\subsubsection{Linear Growth Phase}

\begin{figure}[hbt!]
  \centering
  \includegraphics[]{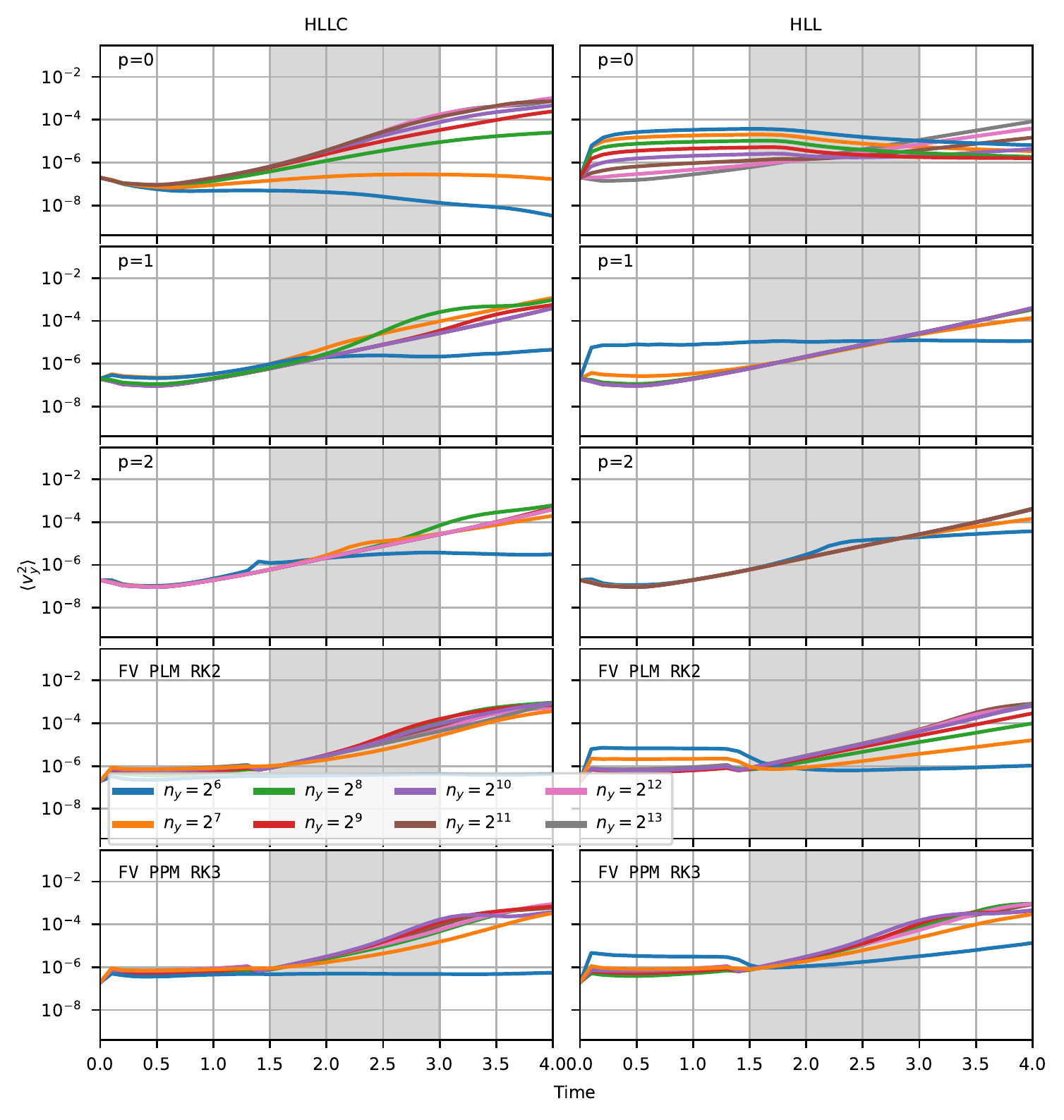}
  \caption{
    \label{fig:kh_mean_squares_vys}
    Mean square of the transverse velocity $v_y$ over time of the relativistic
    2D Kelvin Helmholtz instability using our DG method using a
    $0^{\text{th}}$, $1^{\text{st}}$, and  $2^{\text{nd}}$ order bases
    respectively in the top three rows and using the finite volume code PLUTO
    with PLM and PPM reconstruction respectively in the bottom two
    rows.  In the left column we show results including the contact
    discontinuity in the Riemann solver (using HLLC with our method and HLLD
    with PLUTO) and without  the contact discontinuity using the HLL Riemann
    solver in the right column. The gray band from $t=1.5$ to $t=3.0$ shows the
    region over which we measure the growth rate shown in other plots. Higher
    resolutions generally lead to faster growth rates while the more diffusive
    HLL Riemann solver leads to steadier growth rates due to diminished
    secondary instabilities.
  }
\end{figure}

\noindent We explore the growth of the instability by examining the spatial average
\begin{equation}
  \langle v_y^2  \rangle = \frac{1}{|\Omega|} \int_\Omega v_y^2 \, dV
\end{equation}
where $\Omega$ is the domain and $|\Omega|$ is the volume of the domain. 
Fig.~\ref{fig:kh_mean_squares_vys} shows $\langle v_y^2 \rangle$ as a function
of time in the left column for the Kelvin-Helmholtz instability simulations
explored in this work, where Riemann solvers are grouped by column and basis
order and reconstruction method grouped by rows. Except for the lowest
resolution simulations, all simulations with the HLLC solver enter a linear
growth phase by $t=2.0$ and display non-linear features by $t=4.0$. By
contrast, simulations that utilize the HLL Riemann solver, especially with the
$0^{\text{th}}$ order basis, exhibit large levels of numerical diffusion and
substantially reduced growth rates for all but the largest number of degrees of
freedom. However, for basis order greater than zero, the HLL Riemann solver
exhibits rapid convergence to a well-defined growth rate, while the reference
finite volume schemes that utilize this same Riemann solver exhibit changing
growth rates over this same range of degrees of freedom.

\begin{figure}[hbt!]
  \centering
  \includegraphics[]{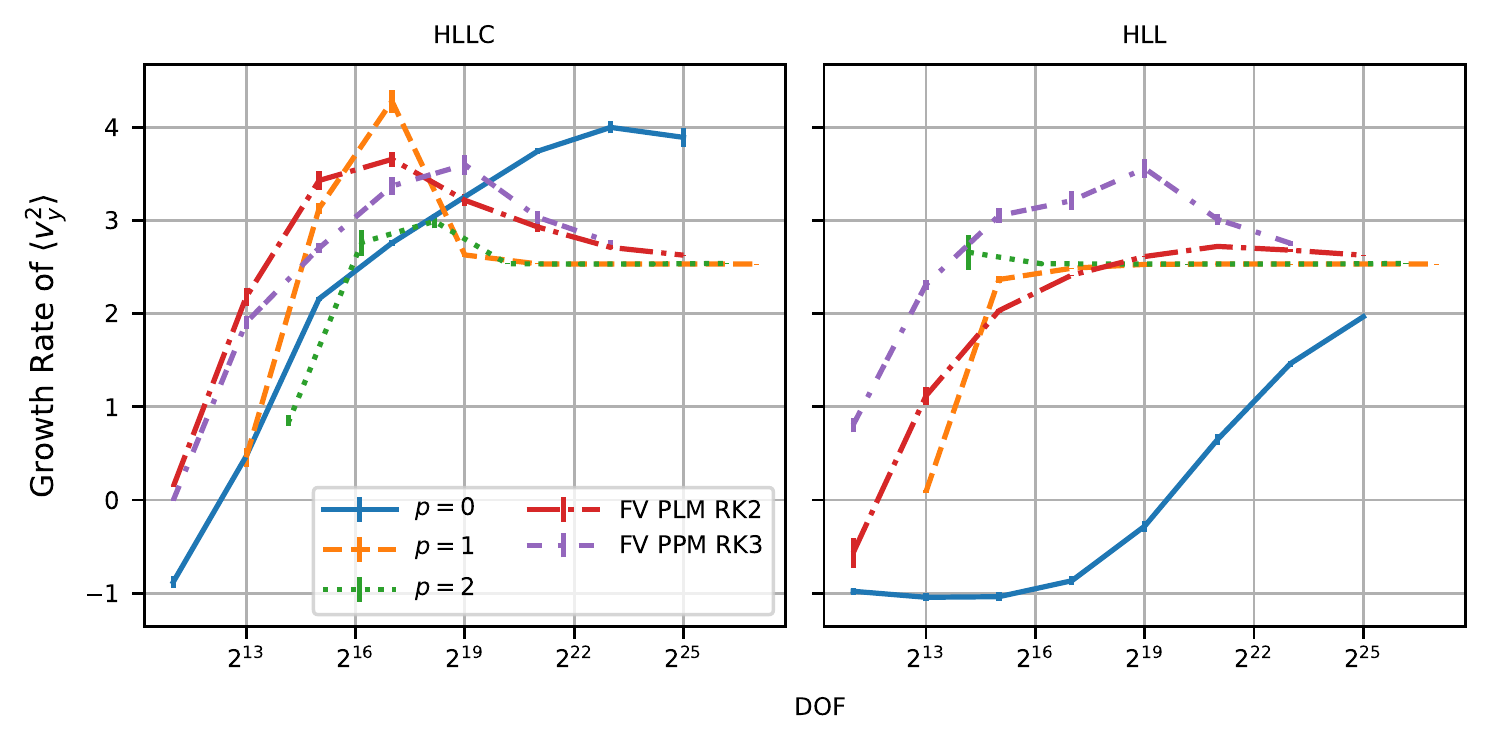}
  \caption{
    \label{fig:kh_growth_rates}
    Growth rates of $\langle v_y^2 \rangle$ versus degrees of freedom from $t=1.5$ to $t=3.0$ of the
    relativistic 2D Kelvin Helmholtz instability using our DG method using the
    finite volume code PLUTO.  In the left column we show results including the
    contact discontinuity in the Riemann solver (using HLLC with our method and
    HLLD with PLUTO) and without  the contact discontinuity using the HLL
    Riemann solver in the right column. Growth rates are measured by computing
    least squares fit of a $\langle v_y^2 \rangle \propto t^\omega$ model to
    the data shown in Fig. \ref{fig:kh_mean_squares_vys}, with error bars
    showing the standard deviation of the least squares fit. 
  }
\end{figure}

We quantify this result by measuring the growth rate, $r$ of $\langle v_y^2 \rangle$
by fitting $\log \langle v_y^2 \rangle(t) = A + r t$ to the measured $\langle v_y^2 \rangle$ using a least squares curve fit in
log space over $t=1.5$ to $t=3.0$. We measure the growth rate early in the
linear growth phase from $t=1.5$ to $t=3.0$ before non-linear modes dominate.
We perform the fit in log space so as to not favor the larger changes in
$\langle v_y^2 \rangle$ at later times. The growth rate of  $\langle v_y^2 \rangle$ for all simulations and methods
versus the degrees of freedom is shown in Fig. \ref{fig:kh_growth_rates}.
Here, the degrees of freedom for a given resolution $n_x \times n_y$ and basis
order $p$ is $\text{DOF} = n_x \times n_y \times (p+1)^2$. Except for the
discontinuous-Galerkin methods using the $0^{\text{th}}$ order basis, the
growth rates using different methods converge to approximately the same value
with higher resolutions.  Generally, using higher order bases, using the HLL
Riemann solver over the HLLC Riemann solver, and using the
discontinuous-Galerkin method over the finite volume method lead to faster
convergence of growth rate. Notably, the overall second order accurate discontinuous-Galerkin
scheme (first order basis, second order time integration scheme) achieves a converged
growth rate at lower numbers of degrees of freedom than a overall second order accurate
finite volume scheme, using either the HLLC or HLL Riemann solver.

\begin{figure*}[hbt!]
  \centering
  \includegraphics[]{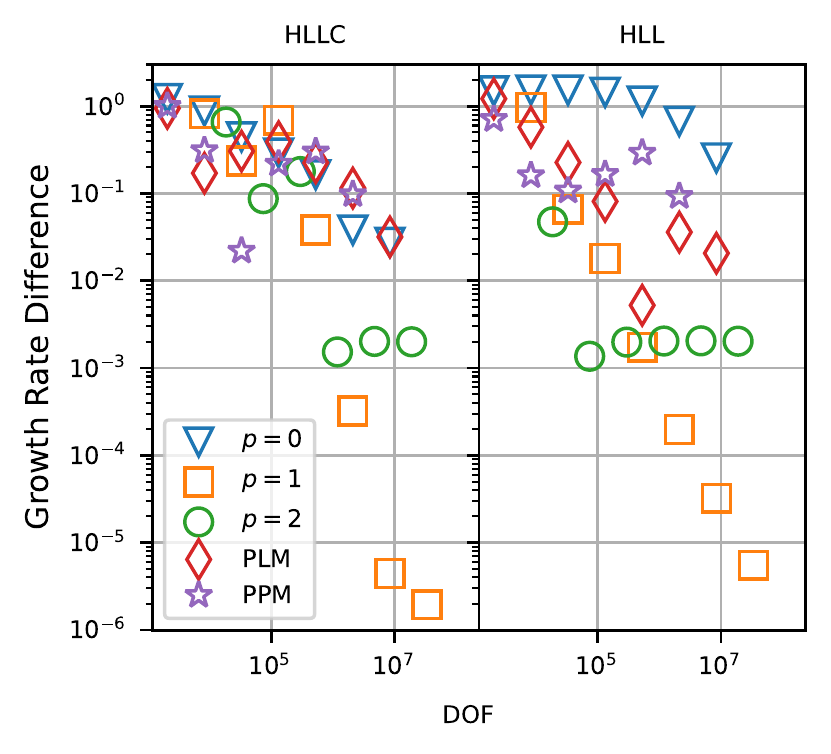}
  \caption{
    \label{fig:kh_growth_rates_vs_dofs}
    The absolute difference in growth rate between the highest resolution
    simulation for each method and each of the lower resolution simulations
    which serves as rough measure of the error of the growth rate, plotted
    versus the degrees of freedom. The discontinuous-Galerkin simulations with
    a $1^{\text{st}}$ order basis show the most effective convergence of the
    simulations explored here, with HLLC converging slightly faster at the
    highest resolutions and HLL converging faster at lower resolutions.  The
    discontinuous-Galerkin simulations with a $2^{\text{nd}}$ order basis do
    not converge below a $10^{-1}$ difference even with high resolutions, which
    we attribute to the boundary effects that worsen with higher resolution.
    Otherwise, the other methods converge at varying rates,  the
    $0^{\text{th}}$ order basis discontinuous-Galerkin methods converging the
    slowest. 
  }
\end{figure*}

This result is explored in more detail in Fig. \ref{fig:kh_growth_rates_vs_dofs}.
The data of this figure shows the difference in growth
rate between the highest resolution simulation with a certain method and the
lower resolution simulations with the same methods versus the degrees of freedom. The
discontinuous-Galerkin simulations with a $1^{\text{st}}$ order basis show the
most effective convergence of the simulations explored here, with HLLC
converging slightly faster at the highest resolutions and HLL converging faster
at lower resolutions. By contrast, the overall second order accurate finite volume
schemes exhibit slower convergence than this scheme, despite the equivalent
order of accuracy, while the first order accurate discontinuous-Galerkin scheme
exhibits similar convergence rates as the finite volume schemes when combined
with the HLLC Riemann solver, but low convergence rates with the HLL solver.
We also note that the discontinuous-Galerkin simulations with a
$2^{\text{nd}}$ order basis do not converge below a $10^{-1}$ difference even
with high resolutions, which we attribute to interaction of the flow with
outflow boundary conditions used here, highlighting the need for improved
fidelity boundary conditions in order to realize the promise of higher order
discontinuous-Galerkin methods.

\subsubsection{Non-linear Evolution}
\label{sec:kh_non_linear}

\begin{figure}
  \centering
  \includegraphics[]{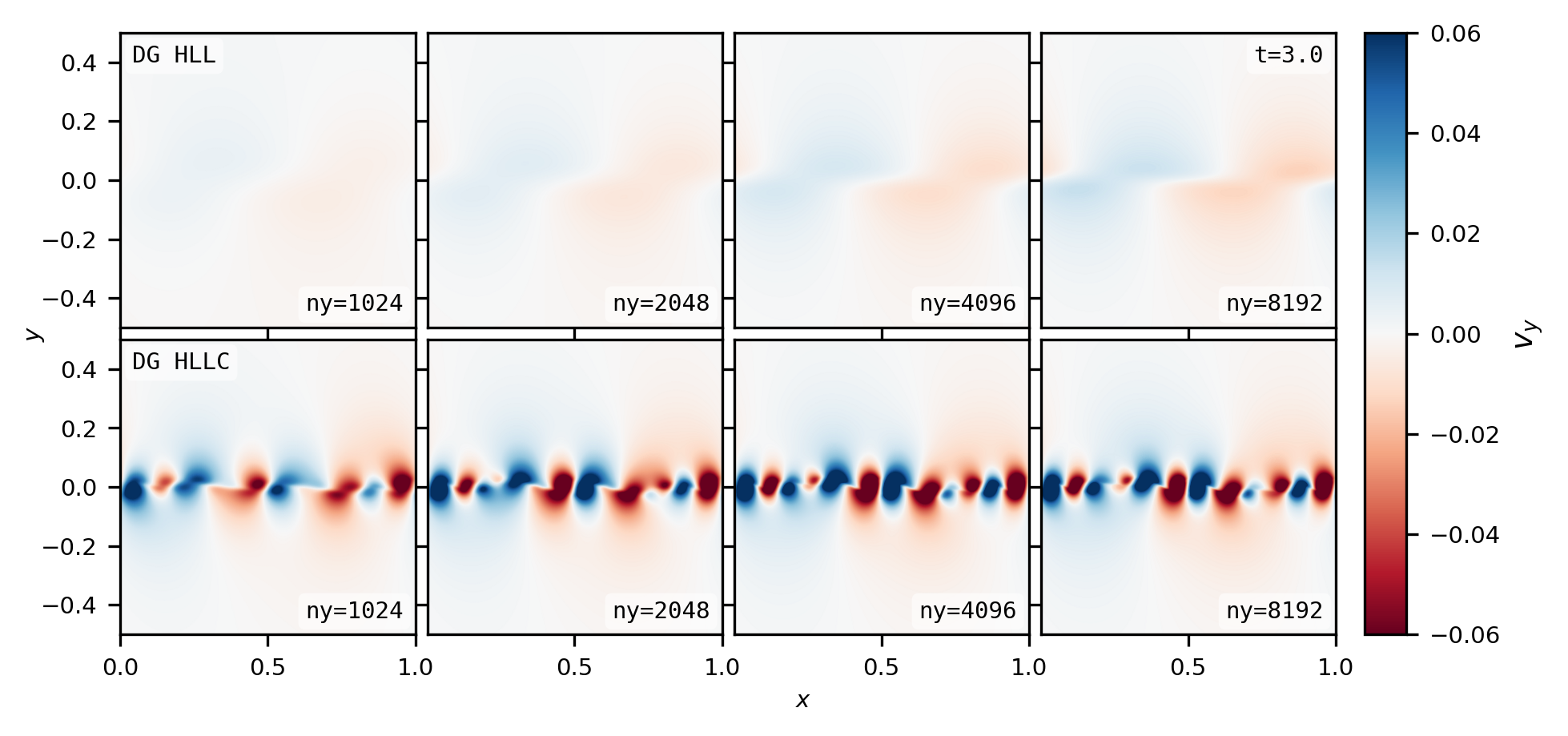}
  \caption{
    \label{fig:kh_vy_30_0}
  Snapshots of the transverse velocity at $t=3.0$ from simulations of the
  relativistic Kelvin-Helmholtz instability using the method presented in this
  work using a $0^{\text{th}}$ order basis. We show results using the HLL
  Riemann solver in the top row and with HLLC in the bottow row. We show the
  four highest resolution simulations across the columns, ranging from
  $512\times1024$ to $4096\times8192$ cells from  left to right. With basis
  order zero, at this stage, using the HLL Riemann solver the method has
  difficulty growing the Kelvin Helmholtz instability, although the structure
  of the perturbation resembles results with simple structures when using
  higher orders. The HLLC Riemann solver generates secondary vortices that get
  worse with high resolutions, which leads to a climbing growth rate.
}
\end{figure}

\begin{figure}
  \centering
  \includegraphics[]{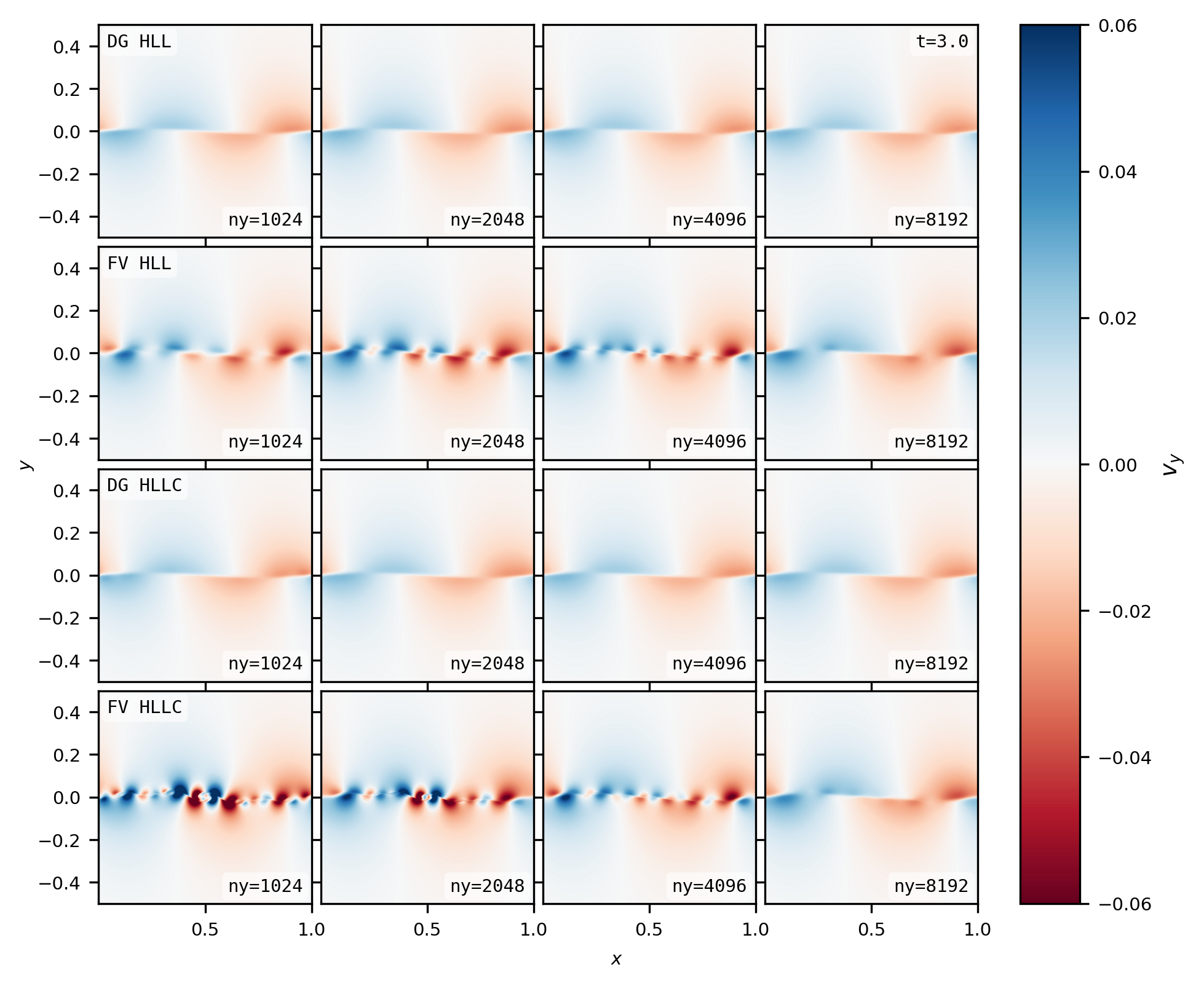}
  \caption{
    \label{fig:kh_vy_30_1}
  Snapshots of the transverse velocity at $t=3.0$ from simulations of the
  relativistic Kelvin-Helmholtz instability using the method presented in this
  work using a $1^{\text{st}}$ order basis in the first and third row and with
  the PLUTO finite volume MHD code with a first order method. We show results
  using the HLL Riemann solver in the top two rows and with HLLC for our code
  and with HLLD for PLUTO in the bottow two rows.  We show the four highest
  resolution simulations across the columns, ranging from $512\times1024$ to
  $4096\times8192$ cells from  left to right. Note that DG method has 4 times
  as many degrees of freedom with the $1^{\text{st}}$ order basis, meaning that
  our $512\times1024$ simulation is comparable in degrees of freedom to the
  $1024\times2048$ simulation using PLUTO. At  this times and these
  resolutions, the results with our DG method have converged to a similar
  solution with a simple structure. Results with PLUTO converge towards the DG
  method results, with secondary vortices present at lower resolutions that
  are more pronounced with HLLC.
}
\end{figure}

\begin{figure}
  \centering
  \includegraphics[]{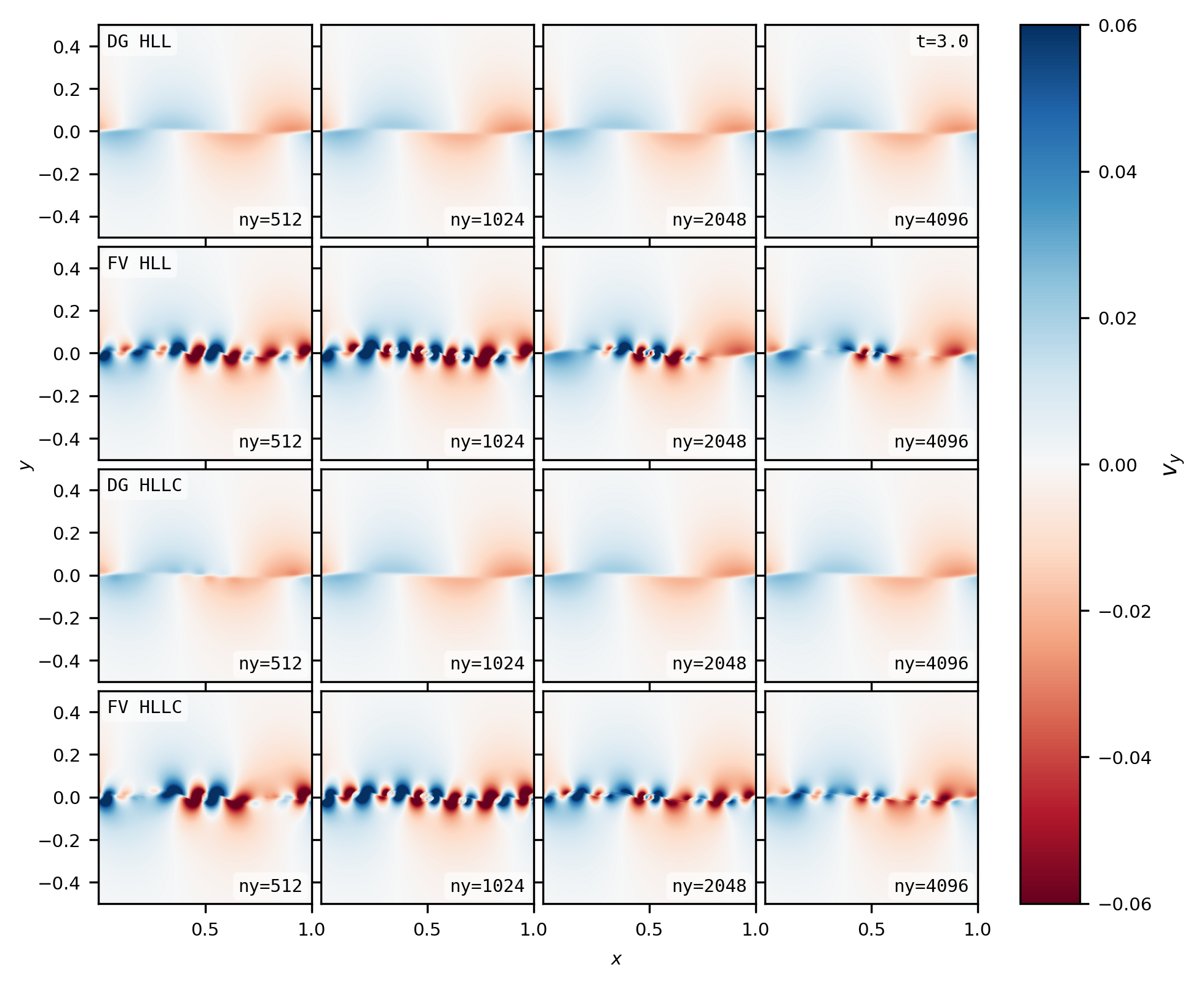}
  \caption{
    \label{fig:kh_vy_30_2}
  Snapshots of the transverse velocity at $t=3.0$ from simulations of the
  relativistic Kelvin-Helmholtz instability using the method presented in this
  work using a $2^{\text{nd}}$ order basis in the first and third row and with
  the PLUTO finite volume MHD code with a second order method. We show results
  using the HLL Riemann solver in the top two rows and with HLLC for our code
  and with HLLD for PLUTO in the bottom two rows.  We show the four highest
  resolution simulations across the columns, ranging from $512\times1024$ to
  $4096\times8192$ cells from  left to right. Note that DG method has 4 times
  as many degrees of freedom with the $1^{\text{st}}$ order basis, meaning that
  our $512\times1024$ simulation has degrees of freedom between the
  $1024\times2048$ simulation and $2048 \times 4096$ simulation using PLUTO.
  With this higher order basis at $t=3.0$, we also see the results with our DG
  method converge quickly to simple structures while the results with PLUTO
  require more resolution to suppress secondary vortices. However, in our
  results using  $4096\times8912$ cells with basis order 2, we see anomalously
  high transverse velocities away from the interface, which is caused by
  boundary effects at high resolutions that will be addressed in future
  improvements to the method.
}
\end{figure}

Fig. \ref{fig:kh_vy_30_0}, \ref{fig:kh_vy_30_1}, and \ref{fig:kh_vy_30_2} show
the state of the Kelvin Helmholtz instability at $t=3.0$ using the method
presented in this work and the reference finite volume scheme \cite{mignonePLUTOCODEADAPTIVE2011} with the 4 highest
resolutions explored in this study. The different figures  show results using
$0^{\text{th}}$, $1^{\text{st}}$, and $2^{\text{nd}}$ order bases or 
$1^{\text{st}}$, $2^{\text{nd}}$, and $3^{\text{rd}}$ order methods
respectively, where a $1^{\text{st}}$ method is only available for our code.
In Fig. \ref{fig:kh_vy_30_0} using our method with a $0^{\text{th}}$ order basis
or a $1^{\text{st}}$ order method, we see significant differences between the
HLL and HLLC solutions; the HLL Riemann solver struggles to
grow the instability, although the structure of the perturbation resembles
results with simple structures when using higher orders. Secondary
instabilities appear to be nonexistent. By contrast, the HLLC Riemann solver generates
secondary vortices that increasing in amplitude with higher resolutions.
Looking at Fig. \ref{fig:kh_vy_30_1} and \ref{fig:kh_vy_30_2}, the
$2^{\text{nd}}$ and $3^{\text{rd}}$ order methods from this work quickly
converge to simple structures. The finite volume method also
converges to a  similar simple structure, although it requires more resolution
compared to the discontinuous-Galerkin method presented here.

\begin{figure}
  \centering
  \includegraphics[]{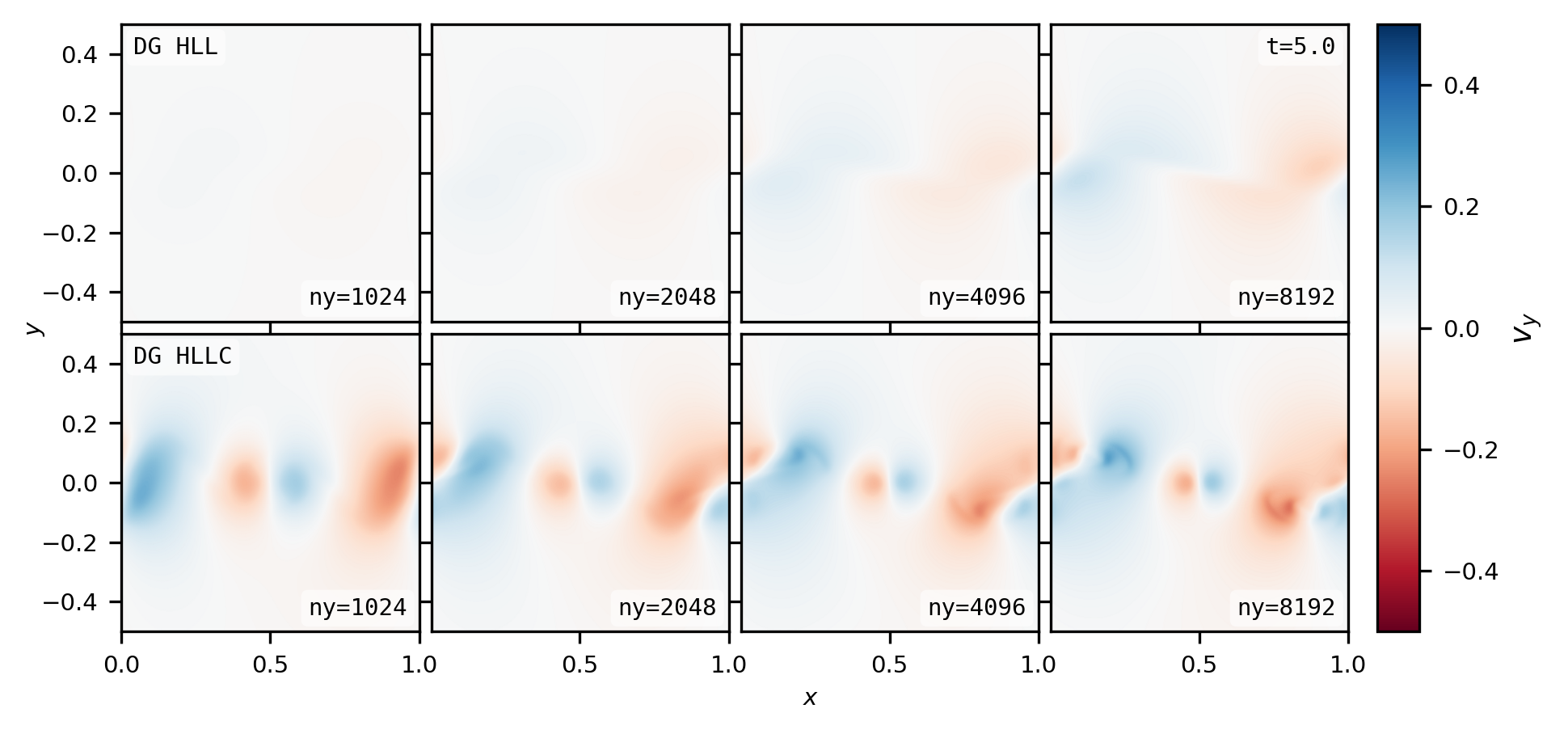}
  \caption{
    \label{fig:kh_vy_50_0}
  Snapshots of the transverse velocity at $t=5.0$ from simulations of the
  relativistic Kelvin-Helmholtz instability using the method presented in this
  work using a $0^{\text{th}}$ order basis. We show results using the HLL
  Riemann solver in the top row and with HLLC in the bottom row. We show the
  four highest resolution simulations across the columns, ranging from
  $512\times1024$ to $4096\times8192$ cells from  left to right. At late times
  into what should be the linear  growth phase, our DG method with the HLL
  solver struggles to growth the instability at low resolutions. The HLLC
  method has developed some structures but they do not resemble results at
  higher orders.
}
\end{figure}

\begin{figure}
  \centering
  \includegraphics[]{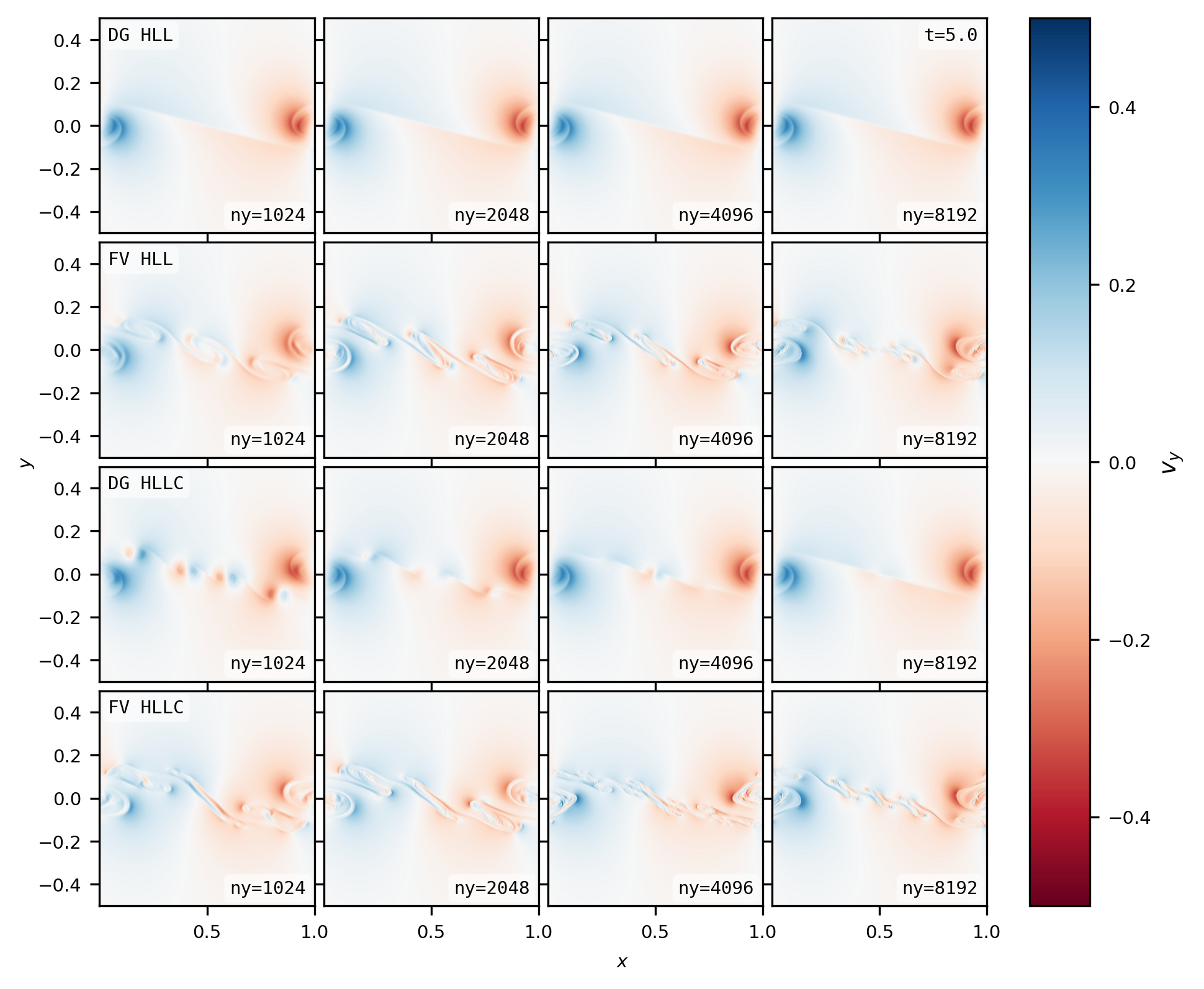}
  \caption{
    \label{fig:kh_vy_50_1}
  Snapshots of the transverse velocity at $t=5.0$ from simulations of the
  relativistic Kelvin-Helmholtz instability using the method presented in this
  work using a $1^{\text{st}}$ order basis in the first and third row and with
  the PLUTO finite volume MHD code with PLM reconstruction. We show results
  using the HLL Riemann solver in the top two rows and with HLLC for our code
  and with HLLD for PLUTO in the bottom two rows.  We show the four highest
  resolution simulations across the columns, ranging from $512\times1024$ to
  $4096\times8192$ cells from  left to right. Note that DG method has 4 times
  as many degrees of freedom with the $1^{\text{st}}$ order basis, meaning that
  our $512\times1024$ simulation is comparable in degrees of freedom to the
  $1024\times2048$ simulation using PLUTO. At this later time once the
  instability has entered into the nonlinear growth phase, the DG method shows
  clear roll ups at all resolutions. Secondary vortices are suppress with
  higher resolutions and by the more diffusive HLL solver. In contrast, the
  PLUTO results show secondary instabilities through out the perturbation,
  although these diminish with resolution. Notably, the structure of the
  instabilities with the DG method versus the finite method are very different.
}
\end{figure}

\begin{figure}
  \centering
  \includegraphics[]{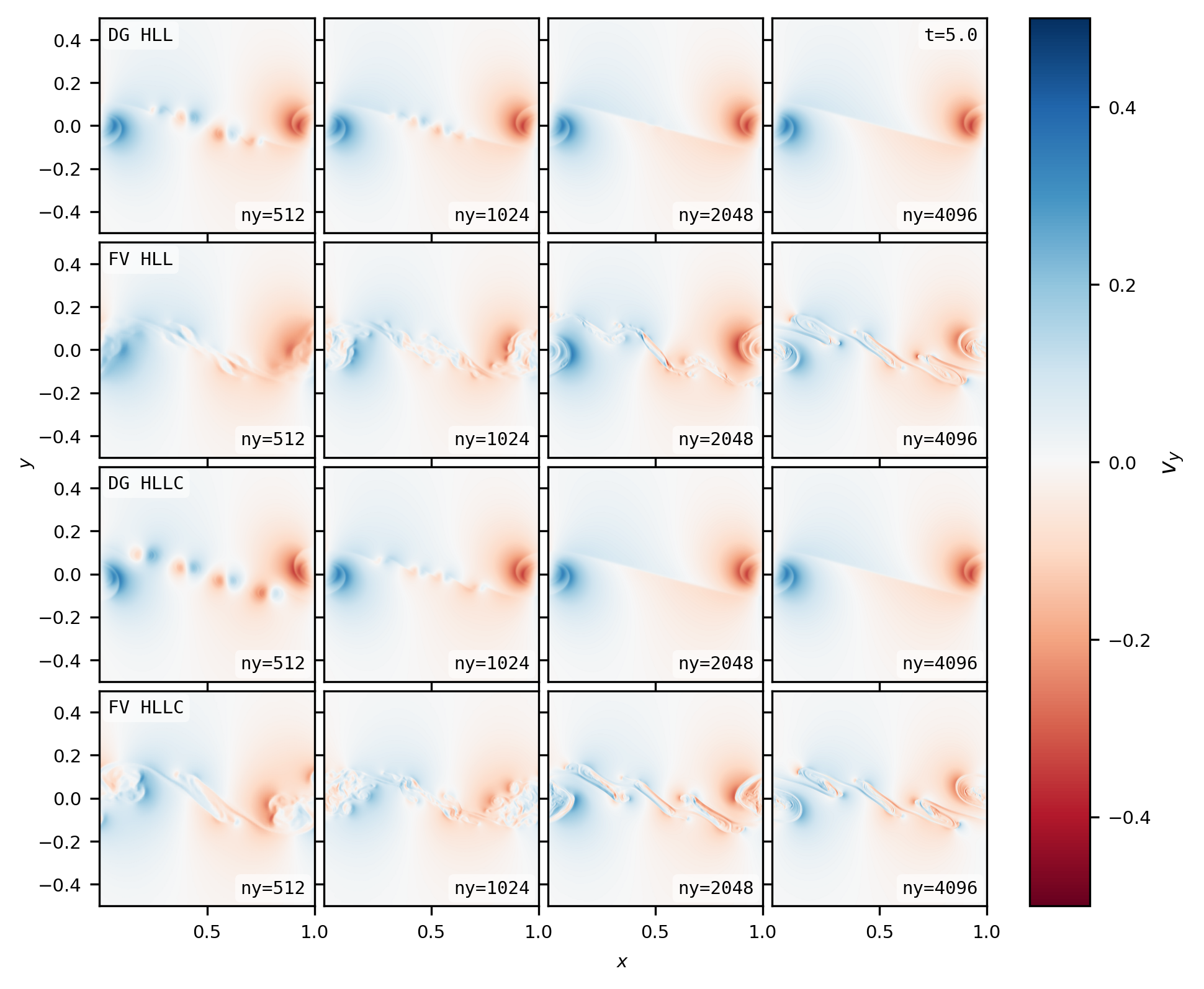}
  \caption{
    \label{fig:kh_vy_50_2}
  Snapshots of the transverse velocity at $t=5.0$ from simulations of the
  relativistic Kelvin-Helmholtz instability using the method presented in this
  work using a $2^{\text{nd}}$ order basis in the first and third row and with
  the PLUTO finite volume MHD code with PPM reconstruction. We show results
  using the HLL Riemann solver in the top two rows and with HLLC for our code
  and with HLLD for PLUTO in the bottom two rows.  We show the four highest
  resolution simulations across the columns, ranging from $512\times1024$ to
  $2048\times4096$ cells from  left to right. Note that DG method has 4 times
  as many degrees of freedom with the $1^{\text{st}}$ order basis, meaning that
  our $512\times1024$ simulation has degrees of freedom between the
  $1024\times2048$ simulation and $2048 \times 4096$ simulation using PLUTO.
  The suppression of secondary vortices with our DG method is enhanced with
  basis order 2 compared to basis order 1, requiring fewer cells and degrees of
  freedom.  Secondary instabilities still appear with the finite volume method,
  largely unaffected by the increase in method order.
}
\end{figure}

Figs. \ref{fig:kh_vy_50_0}, \ref{fig:kh_vy_50_1}, and  \ref{fig:kh_vy_50_2} show
the state of the Kelvin Helmholtz instability at $t=5.0$, which is well into
the non-linear phase, using the method presented in this work and with the
reference finite volume scheme with the 4 highest resolutions explored in this study.
The different figures show results using $1^{\text{st}}$, $2^{\text{nd}}$, and
$3^{\text{rd}}$ order methods respectively, where a $1^{\text{st}}$ is only
available for our code.
In Fig. \ref{fig:kh_vy_50_0} using our method with a $0^{\text{th}}$ order basis
or a $1^{\text{st}}$ order method, we again see significant differences between
the HLL and HLLC solutions. The HLLC solution grows faster than the HLL
solution but neither resemble the structures seen with higher order bases.
Using the HLLC Riemann solver, secondary vortices are apparent during the
non-linear phase, which become more defined with higher resolution.
Examining Figs. \ref{fig:kh_vy_50_1} and \ref{fig:kh_vy_50_2}, the $2^{\text{nd}}$
and $3^{\text{rd}}$ order methods from this work quickly converge with higher
resolution to simple structures during the non-linear phase. Results with HLL
over HLLC and with a $2^{\text{nd}}$ order basis over a $1^{\text{st}}$ order
basis are generally smoother with fewer secondary vortices. The solution generated
by the reference finite volume scheme also converges to roughly the same
structures as the discontinuous-Galerkin method, although secondary
instabilities are obvious along the interface between the primary vortices.
Note that the mode of these secondary instabilities increased with resolution,
with smaller but more numerous instabilities at higher resolutions.

Our interpretation of these results is that the secondary structures found in
the finite volume method at the end of the linear growth phase serve to seed
non-linear structures that are observed at late times; a result somewhat
consistent with that reported by \cite{Lecoanet:2016}.  What is notable is that
these structures vanish in the second order accurate (first order basis)
discontinuous-Galerkin scheme presented here at lower resolution than in the
finite volume scheme for the HLLC Riemann solver and are \emph{absent} in the
HLL Riemann solver based scheme, indicating a role played by the dissipation of
the HLLC Riemann solver in the formation of these structures. In addition, the
presence of these structures in the finite volume scheme utilizing the HLL
solver and the clear dependency of the properties of these structures on the
reconstruction method (PLM vs. PPM) is another point of contrast between
discontinuous-Galerkin methods and finite volume schemes.  This is strongly
reminiscent of the results presented by \cite{Lecoanet:2016}, where finite
volume schemes were demonstrated to exhibit similar secondary vortices at
moderate resolutions (similar to these presented here), which then disappeared
at higher resolutions; the higher resolution simulations being comparable to
spectral methods. The absence of such secondary vortices for combinations of
the discontinuous-Galerkin algorithms presented here suggest that these methods
may be less susceptible to such considerations.  However, we stress that this
is a single application on both methods and that the performance of either
method may depend on details of the set up of the instability.  Over the
development of the method, we also explored the analytic growth rate of
perturbations given the initial conditions from
\citet{bodoKelvinHelmholtzInstabilityRelativistic2004}, where we found that the
growth rate of the instability generally did not match the analytically
predicted growth rate, and that an initial transient outgoing wave from the
initial perturbation caused significant boundary effects with the
$2^{\text{nd}}$ order basis. Although our discontinuous-Galerkin method
provides apparently better results in this case, more development especially
around boundary conditions is required.

\subsection{Performance} \label{sec:performance}

To test the performance of the method on multiple architectures, we timed
simulations of the Kelvin Helmholtz instability on CPUs and GPUs, using the
perturbations described in \S\ref{sec:kelvin_helmholtz}. For both
architectures, we time the performance of the code with $v_{x,0}=0.25c$ using
basis orders 0, 1, and 2 and resolutions of $256\times512$, $512\times1024$,
and $1024\times2048$ with each basis order testing both HLLC and HLL for a
total of 18 simulations for both architectures. 
We conduct CPU testing on  1024 cores spread across 22 dual socket nodes with
Intel Xeon Platinum 8268 CPUs, comprising approximately $\sim 88
\text{TFLOPS}$ in total. For GPU runs we use 32 NVidia Tesla V100-SXM2 GPUs
spread across 8 nodes, comprising approximately $\sim 250 \text{TFLOPS}$ in
total.  These computational resources were chosen to accommodate the memory
needed for the largest simulation in the performance profiling suite.

\begin{figure}[hbt!]
  \centering
  \includegraphics[]{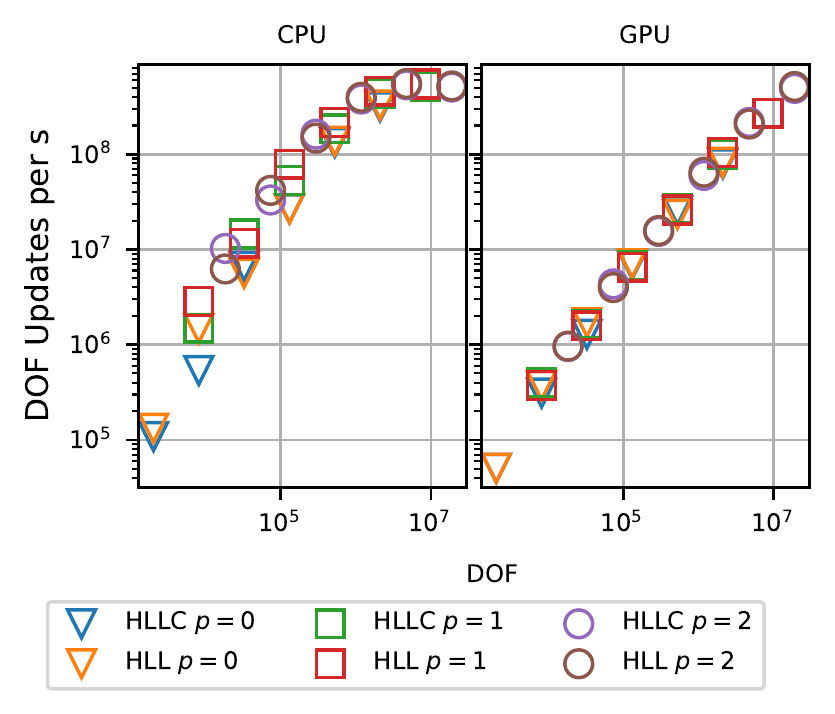}
  \caption{
    \label{fig:kh_performance} Performance of the code modeling the Kelvin
    Helmholtz instability from section \S\ref{sec:kelvin_helmholtz}, plotting
    updates to degrees of freedom per second versus degrees of freedom, using
    1024 cores spread across 22 dual socket nodes with Intel Xeon Platinum 8268
    CPUs (comprising approximately $\sim 88 \text{TFLOPS}$ in total) in the
    left column and using 32 NVidia Tesla V100-SXM2 GPUs (comprising
    approximately $\sim 250 \text{TFLOPS}$ in total) spread across 8 nodes on
    the right, where the peak computational throughput of the GPUs used are
    roughly three times the peak computational throughput of the CPUs. The
    computational resources for both tests was chosen to accommodate the memory
    needed for the largest simulation in the suite.  We show profiling results
    with the HLLC and HLL Riemann solvers and  with the $0^{\text{th}}$,
    $1^{\text{st}}$, and $2^{\text{nd}}$ order bases, between which we see
    little difference in performance. Comparing between the CPU and GPU runs,
    however, we see that the CPU performance becomes saturated at around $10^6
    \text{ DOFs}$ while the GPUs have not saturated the performance, even with
    simulations using more than $10$ times the degrees of freedom.
  }
\end{figure}

We show profiling results with the HLLC and HLL Riemann solvers and  with the
$0^{\text{th}}$, $1^{\text{st}}$, and $2^{\text{nd}}$ order bases in Fig.
\ref{fig:kh_performance}. The degree of freedom updates per second is computed
with
\begin{equation}
  \text{DOF per second} = \frac{ \text{DOF} \times \text{steps} 
    \times \text{stages per step}  }
  {\text{ time to solution in seconds} },
\end{equation}
which serves as a measure of computational efficiency.  With the \texttt{RK1},
\texttt{SSPRK2}, and \texttt{SSPRK3} integrators used for basis orders $0$,
$1$, and $2$ we use $1$, $2$, and $3$ stages per step for the respective basis
orders.     

We show profiling results with the HLLC and HLL Riemann solvers and  with the
$0^{\text{th}}$, $1^{\text{st}}$, and $2^{\text{nd}}$ order bases, between
which we see little difference in performance.  Comparing between the CPU and
GPU runs,  we see that the CPU performance becomes saturated at around $10^6
\text{DOF}$ while the GPUs have not saturated the performance, even with
simulations using more than $10$ times the degrees of freedom.  Simulations
with more degrees of freedom  would not fit within GPU memory here, indicating
that our present implementation is unable to fully saturate GPU performance.
Note that the theoretical peak throughput of the GPU resources using here is
approximately three times the throughput for the CPU resources. Memory
bandwidth resources between RAM and the registers on CPUs and HBM memory and
the registers on GPUs is similarly greater on GPUs Since the CPUs and GPUs
achieve roughly the same updates per second, this indicates underutilization of
GPU FLOPS. i.e. our implementation is failing to meet computation or memory
bounds, where the arithmetic-intensity of discontinuous-Galerkin methods lead
to typically memory bound algorithms.

These performance characteristics are consistent with insufficient work within
individual kernels to offset kernel launch overhead, as was the case in the
\kathena magnetohydrodynamics code presented in
\citet{greteKAthenaPerformancePortable2021} and was resolved in the \Parthenon
adaptive-mesh refinement framework and \AthenaPK magnetohydrodynamics code
presented in \citet{greteParthenonPerformancePortable2022}.  We performed an
informal profiling of our method evolving the Kelvin-Helmholtz instability on a
single V100 GPU  using \texttt{nvprof}. With a timeline trace, we verified for
problem sizes that occupied the entirety of the HBM memory of a single GPU that
a large percentage of compute time on the GPU, $>70\%$, was dominated by short
duration $4 \upmu\text{s}$ kernel calls. These kernel durations would be
consumed by kernel launch overhead from within the CUDA API.

 With the launch of each kernel, between the APIs, drivers, and hardware a few
 microseconds are spent launching the kernel on the GPU. Unless sufficient work
 is done within each kernel, this launch overhead will dominate runtime. For
 our implementation, the work done within individual kernels can be increased
 with more degrees of freedom.  However, the GPU has insufficient memory to
 allow enough work to hide kernel launch overhead, hence the underutilization
 of the GPU. In \Parthenon and \AthenaPK, this kernel launch overhead was
 hidden by fusing together the work from multiple kernels into fewer, larger
 kernels\cite{greteKAthenaPerformancePortable2021}.  Similar improvements would
 be needed for our implementation in order to saturate GPU performance.

\section{Summary} \label{sec:conclusions}
We have presented a scheme to evolve the relativistic hydrodynamics equations
using a discontinuous-Galerkin method.  Within our scheme, we have developed a
robust method for enforcing physicality of the conserved state via a operator.
Our presentation of the method includes relativistic HLL and HLLC Riemann
solvers,  multiple methods for recovering the primitive variables from
conserved variables with the ideal equation of state, and the Taub-Matthews
approximation to the Synge equation of state, using physical units that keep
factors of $c$. We implement the method using the Kokkos performance
portability library, which allows us to run CPUs and GPUs supported by Kokkos.

The novel physicality-enforcing operator in the work allows evolution of shocks
with high-order basis methods. The operator strictly enforces positive density
and pressure and subluminal velocities on all basis points within a cell by
smoothing nonphysical points towards the physical volume average. Additionally,
the method conserves volume averages of conserved variables.

In our exploration of methods to recover primitive variables from conserved
variables when using an ideal equation of state, we found that the iterative
method from \citet{riccardiPrimitiveVariableRecovering2008} was faster, more
robust, and more accurate than the analytical method from
\citet{ryuEquationStateNumerical2006}, consistent with findings from
\citet{riccardiPrimitiveVariableRecovering2008}. The iterative method for ideal
gases presented here recovers the primitive variables by solving a quartic as
described in Eq.~\ref{eq:iterative_quartic_unknown}, which provides more digits
of precision in simultaneously in sub-relativistic and ultra-relativistic
regimes compared to solving in terms of the velocity or Lorentz factor.
Additionally, the Newton-Raphson method as applied to
Eq.~\ref{eq:iterative_quartic} gives comparable accuracy to the analytic method
in under 10 iterations, as is explored in
Fig.~\ref{fig:conserved_to_primitive_accuracy}. More iterations allow a more
accurate recovery with the iterative method compared to the analytic method.
In the case of our implementation, the iterative method is faster to compute
for $\gamma < 10$ on CPUs and always faster on GPUs except in trivial cases.

Conversely, in our exploration of methods to recover primitives variables from
conserved variables with the Taub-Mathews equation of state, the analytical
method detailed in \citet{ryuEquationStateNumerical2006} was faster than the
iterative method implemented in this work. With the Taub-Mathews equation of
state, recovering the primitives requires solving a cubic equation, which has a
much simpler analytical solution compared to the quartic equation for the ideal
gas. Solving this cubic equation iteratively requires a bounded root solver,
where we use Brent's method in this work. The iterative method we implemented
for the Taub-Matthews equation of state requires many more iterations to
achieve acceptable accuracy than the iterative solver for the ideal gas. As
such, we found the analytic method for the Taub-Matthews equation of state to
outperform the iterative method in terms of time to solution and accuracy on
both CPUs and GPUs.

With this method, we ran several standard test problems, including linear
waves, 1D and 2D Riemann problems, and the relativistic Kelvin-Helmholtz
problem. The iterative conserved-to-primitive solver facilitated more
relativistic problems and the physicality-enforcing operator allowed stable
evolution with higher order bases for problems with shocks. In some test
problems with a shock moving transverse to an outflow boundary conditions, we
saw some non-physical boundary effects when using a $2^{\text{nd}}$ order basis.

In our tests of the Kelvin-Helmholtz instability, comparing to results using
a finite volume reference scheme \citep{mignonePLUTOCODEADAPTIVE2011}, the
discontinuous-Galerkin method presented in this work can better suppress
secondary vortices and instabilities compared to the finite volume method. Our
method works best with a $1^{\text{st}}$ order basis, which is a
$2^{\text{nd}}$ order method in space and time, since the $0^{\text{th}}$ order
basis is slow to grow the instability with low resolution while with the
$2^{\text{nd}}$ order basis boundary effects enter in at the outflow boundaries
with high resolution. 

In the tests of the Kelvin-Helmholtz instability and some of the 2D Riemann
problems, we saw numerical boundary effects enter at the outflow boundary
conditions, which increased with higher resolutions. Further development of the
outflow boundaries with higher order bases is required.

Finally, in the exploration of the performance of our implementation evolving
the Kelvin-Helmholtz instability, we found that our implementation is unable to
saturate performance on GPUs before the problem size grows too large for the
GPU memory. From these performance results and profiling using \texttt{nvprof},
we suspect that insufficient work inside individual kernels, leading to kernel
launch overhead dominating runtime, is responsible for the lack of performance
on GPUs. Combining the work from multiple kernels -- as was done in the
\Parthenon framework presented in \citet{greteKAthenaPerformancePortable2021}
-- would be needed for our implementation in order to saturate GPU performance.

\section*{Acknowledgments} \label{sec:acknowledgements}

Sandia National Laboratories is a multimission laboratory managed and operated
by National Technology \& Engineering Solutions of Sandia, LLC, a wholly owned
subsidiary of Honeywell International Inc., for the U.S. Department of Energy's
National Nuclear Security Administration under contract DE-NA0003525. This
paper describes objective technical results and analysis. Any subjective views
or opinions that might be expressed in the paper do not necessarily represent
the views of the U.S. Department of Energy or the United States Government.
This work was supported in part by LDRD project \#209240. SAND~\#SAND2022-3784~O

\end{document}